\documentclass[twocolumn, twocolappendix]{aastex631}
\usepackage{float}
\usepackage{wasysym}
\usepackage{booktabs}
\usepackage{changepage}
\usepackage[encapsulated]{CJK}

\begin{document}
\begin{abstract}
Early JWST studies found an apparent population of massive, compact galaxies at
redshifts $z\gtrsim7$. Recently three of these galaxies were shown to have prominent
Balmer breaks, demonstrating that their light at $\lambda_{\rm rest} \sim 3500$\,\AA\
is dominated by a stellar population that is relatively old ($\sim$200 Myr). All three also have broad H$\beta$ emission with
$\sigma > 1000 \,\rm km \, s^{-1}$, a common feature of such
`little red dots'.
From S\'ersic profile fits to the NIRCam images in F200W we find that the stellar light of galaxies is extremely compact: the galaxies have half-light radii of $r_{\rm e}\sim$  100 pc, in the regime of ultra compact dwarfs in the nearby
Universe.
Their masses are uncertain, as they depend on the contribution of possible
light from an AGN to the flux at $\lambda_{\rm rest}>5000$\,\AA. If the AGN contribution
is low beyond the Balmer break region,
the masses are $M_* \sim 10^{10}-10^{11}\mathrm{M}_{\astrosun}$, and the central densities 
are higher than those of any other known
galaxy population by an order of magnitude. Interestingly, the
implied velocity dispersions of $\sim$1500 km s$^{-1}$ are in very good
agreement with the measured H$\beta$ line widths. 
We suggest that some of the broad lines in `little red dots' are not due to AGNs but simply reflect the kinematics of the galaxies,
and speculate that the galaxies are observed in a short-lived phase where the central densities are much higher than at later times. We stress, however, that
the canonical interpretation of AGNs causing
the broad H$\beta$ lines also remains viable.
\end{abstract}

\title{The Small Sizes and High Implied Densities of `Little Red Dots' with Balmer Breaks Could Explain Their Broad Emission Lines Without an AGN}
\author[0009-0005-2295-7246]{Josephine F.W. Baggen}
\correspondingauthor{Josephine F.W. Baggen}
\email{josephine.baggen@yale.edu}
\affiliation{Department of Astronomy, Yale University, New Haven, CT 06511, USA}
\author[0000-0002-8282-9888]{Pieter van Dokkum}
\affiliation{Department of Astronomy, Yale University, New Haven, CT 06511, USA}
\author[0000-0003-2680-005X]{Gabriel Brammer}
\affiliation{Cosmic Dawn Center (DAWN), Niels Bohr Institute, University of Copenhagen, Jagtvej 128,
K{\o}benhavn N, DK-2200, Denmark}
\author[0000-0002-2380-9801]{Anna de Graaff}
\affiliation{Max-Planck-Institut f\"ur Astronomie, K\"onigstuhl 17, D-69117, Heidelberg, Germany}

\author[0000-0002-8871-3026]{Marijn Franx}
\affiliation{Leiden	Observatory,	
Leiden	University,	P.O.	Box	9513,	NL-2300	RA	Leiden,	Netherlands}

\author[0000-0002-5612-3427]{Jenny Greene}
\affiliation{Department of Astrophysical Sciences, Princeton University, 4 Ivy Lane, Princeton, NJ 08544, USA}
\author[0000-0002-2057-5376]{Ivo Labb\'e}
\affiliation{Centre for Astrophysics and Supercomputing, Swinburne University of Technology,
Melbourne, VIC 3122, Australia}
\author[0000-0001-6755-1315]{Joel Leja}
\affiliation{Department of Astronomy \& Astrophysics, The Pennsylvania State University, University Park, PA 16802, USA}
\affiliation{Institute for Computational \& Data Sciences, The Pennsylvania State University, University Park, PA 16802, USA}
\affiliation{Institute for Gravitation and the Cosmos, The Pennsylvania State University, University Park, PA 16802, USA}
\author[0000-0003-0695-4414]{Michael V. Maseda}
\affiliation{Department of Astronomy, University of Wisconsin-Madison, 475 N. Charter St., Madison, WI 53706, USA}
\author[0000-0002-7524-374X]{Erica J. Nelson}
\affiliation{Department for Astrophysical and Planetary Science, University of Colorado, Boulder, CO, USA}
\author[0000-0003-4996-9069]{Hans-Walter Rix}
\affiliation{Max-Planck-Institut f\"ur Astronomie, K\"onigstuhl 17, D-69117, Heidelberg, Germany}
\author[0000-0001-9269-5046]{Bingjie Wang (\begin{CJK*}{UTF8}{gbsn}王冰洁\ignorespacesafterend\end{CJK*})}
\affiliation{Department of Astronomy \& Astrophysics, The Pennsylvania State University, University Park, PA 16802, USA}
\affiliation{Institute for Computational \& Data Sciences, The Pennsylvania State University, University Park, PA 16802, USA}
\affiliation{Institute for Gravitation and the Cosmos, The Pennsylvania State University, University Park, PA 16802, USA}
\author[0000-0001-8928-4465]{Andrea Weibel}
\affiliation{Department of Astronomy, University of Geneva, Chemin Pegasi 51, 1290 Versoix, Switzerland}

\section{Introduction}
The James Webb Space Telescope (JWST) has revealed a population of 
 high-redshift objects with very specific spectral energy distributions (SEDs): a flat blue continuum and a steep red slope \citep[e.g.,][]{Labbe2023, Barro2024, Akins2024, Kocevski2023, Ono2022, Onoue2023, Labbe2023b}. 
They are often compact \citep{Akins2023, Baggen2023} \citep[appearing as `little red dots', LRDs;][]{Matthee2024} and some, but not all, are confirmed to have moderately broad (FWHM $\sim$1200-4000 km s$^{-1}$) Balmer lines in their spectra \citep[e.g.,][]{Kocevski2023, Matthee2024, Greene2024, Kokorev2023, Killi2024, Maiolino2024_jades, Ubler2023, Harikane2023, Larson2023, Wang2024a, Wang2024_balmer, Kokorev2024, Kocevski2024}.
The most straightforward interpretation of the small sizes and
broad Balmer lines is that we
are seeing active galactic nuclei (AGN) with a broad line region (BLR).
If they are AGN, they are unusual: they are typically not detected in X-ray emission \citep{Ananna2024,Akins2024, Maiolino2024,  Yue2024, Furtak2024, Greene2024, Kokubo2024}, have yet to demonstrate photometric variability in the Near Infrared Camera \citep[NIRCam;][]{Kokubo2024}, have flat SEDs in the mid-far IR for some LRDs implying dominant contribution from stars at these wavelengths \citep{PerezGonzalez2024A}, show no dominant contribution from hot dust emission expected from an AGN at rest-frame 3 $\mu$m \citep{Williams2024}, and
have not been detected in stacks of mid-IR/far-IR/sub-mm/radio data \citep{Akins2024}.

The correct interpretation of these galaxies -- whether they are dominated by AGNs or are compact and massive -- has far-reaching implications for black hole physics \citep[e.g.][]{Greene2024, Maiolino2024, Bogdan2024, Kovacs2024, Juodzbalis2024, Inayoshi2024}, the theory of galaxy formation and evolution \citep[e.g.][]{Silk2024, Loeb2024}, and perhaps even cosmology \citep{BoylanKolchin2023}.

Recently, three LRDs at redshifts $z = 6.8-8.4$ were shown to have prominent Balmer breaks in their JWST spectra \citep{Wang2024_balmer}. Two of these galaxies are among the massive galaxy candidates reported in \citet{Labbe2023}. 
The Balmer break indicates the presence of an evolved stellar population, with stellar ages of more than 100 Myr \citep{Bruzual1983, Hamilton1985, Worthey1994,Balogh1999, Poggianti1999}, and their detection strongly suggests that
the SEDs at $\lambda_{\rm rest} \sim 3500$\,\AA\ are
not dominated by AGNs. However, all three objects also have broad H$\beta$ emission lines ($\sigma_{\rm H\beta} >$1000 km s$^{-1}$), suggesting that AGNs are present.
\citet{Wang2024_balmer} fit the
SEDs of the three galaxies, and show that the derived
stellar population parameters 
depend sensitively on the interpretation of the continuum redward of the Balmer break.
Equally good fits are obtained using either a model with a high stellar mass and extended SFH, a model with a dust-obscured AGN with minimal stellar mass, or a solution in between. This leads to a wide range of stellar masses ($M_*\sim10^{9}-10^{11}\mathrm{M}_{\astrosun}$) and the implied possible natures and evolutionary tracks of these sources range from the progenitors of massive ellipticals to low-mass galaxies with over-massive black holes. 

In this Letter, we use the fact that we know both the
redshifts of these galaxies and that the extended light in the UV suggests that the light at
these wavelengths is dominated by stars
to measure accurate half-light radii. This follows
our initial size measurements \citep{Baggen2023} of the \citet{Labbe2023}
galaxies that were based on photometric
redshifts and earlier versions of the imaging data.
We then use the previously-determined masses for
different AGN contributions, combined with the sizes,
to measure densities and predict emission line widths and compare
these to the observed line widths.
Throughout this work we assume $\Lambda$CDM cosmology with $H_0$=70 km s$^{-1}$ Mpc$^{-1}$, $\Omega_{m,0}$=0.3 and $\Omega_{\Lambda,0}$=0.7. Magnitudes are reported in the AB system.

\section{Data}
\label{sec:data}
\subsection{Imaging}
\label{sec:imaging}
\begin{figure}[htp!]
    \centering
    \includegraphics[width=0.9\linewidth]{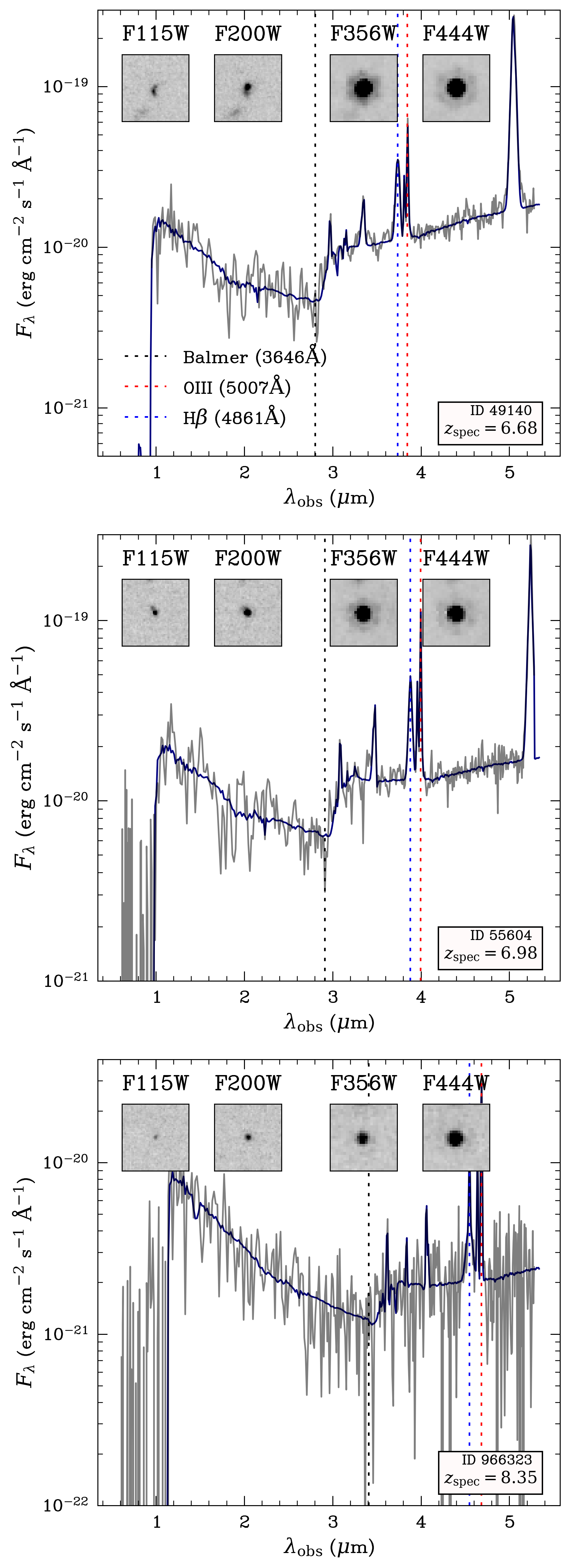}
    \caption{The spectra (grey) and best-fit model (darkblue) with maximal stellar mass of three Balmer break galaxies, obtained from \citet{Wang2024_balmer}. We also show $1\arcsec \times 1\arcsec$ images in four different bands, F115W, F200W, F356W and F444W. The dotted lines indicate the location of the Balmer break (black), [OIII] line (red), and H$\beta$ (blue) line, which are relevant in this work. For information on other emission lines, see \citet{Wang2024_balmer}.  }
    \label{fig:spectra}
\end{figure}

The JWST
Near Infrared Camera (NIRCam) imaging data of this Letter are obtained from the Cosmic Evolution Early Release Science (CEERS) program \citep[PI: Finkelstein; PID: 1345][]{Finkelstein2022, Finkelstein2023} \citep[see data DOI:][]{doiCEERSmosaic}. In this work, we use the mosaics available through the DAWN JWST Archive (DJA, v7.2), which are reduced using the \textsc{grizli} pipeline \citep{gabe_brammer_grizli}. 
The mosaics are available in six broadband filters, short-wavelength (SW; F115W, F150W, F200W) and long-wavelength 
 (LW; F277W, F356W and F444W) with a resolution of $0\farcs 04$\,pix$^{-1}$. In addition, higher resolution ($0\farcs 02$\,pix$^{-1}$) data are available for the SW bands, for which the mosaics are split into 12 tiles (ceers-full-xi.yj-v7.2-f200w-clear\_drc\_sci.fits.gz, where i runs from 0--5 and j from 0--1). 
 Ideally, we would use the band at the location of the Balmer break ($\sim3\mu$m), to do the analysis. However, due to the large pixel size (and the broader PSF) of the LW data, the galaxies are barely resolved/unresolved at these wavelengths. Therefore, we use F200W for our analysis, the reddest band available with 20\,mas sampling. Moreover, the F200W filter probes the highest SNR out of all the SW filters. We show cutouts and RGB images (see Figure \ref{fig:spectra}) of the galaxies using multiple bands for illustration.

Compared to previous versions \citep[e.g. v4 used in][]{Labbe2023, Baggen2023}, the weighting scheme to do the drizzling has changed. Previously, the weighting was done including the Poisson noise from the source. This caused a suppression of flux in the central pixels for compact sources, which could directly affect the sizes of sources in the images. As we show in this work, the two sources (L23-38094, L23-14924) for which the sizes are also reported in \citet{Baggen2023}, the
newly measured sizes are indeed smaller
in the new reductions.

The form of the
point-spread function (PSF), which describes how point sources appear in the data, is a crucial aspect of the
analysis, yet complicated to model properly. 
One approach is to create an empirical PSF from either bright well-centered single stars or by stacking multiple stars in the mosaic. 
Another approach is forward modeling of the telescope instruments, as is done for the WebbPSF tool created for JWST imaging \citep{Perrin2014}. The advantage of these synthetic PSFs is the high signal-to-noise ratios and perfect sampling. However, a variety of studies have found discrepancies between WebbPSF profiles and point sources in the images \citep{Ding2022, Ono2022, Onoue2023, Weaver2024}. 

We first consider a PSF made from stacking of centered, bright, unsaturated stars in the entire mosaic \citep[as described in][]{Weibel2024}. 
However, we find that this PSF gives large ellipticities (and a very similar position angle) when fit to compact
galaxies. Furthermore, the  light profile is clearly
broader than that of single, well-centered stars in the vicinity of the three galaxies. Therefore we also consider five single stars (RA [214.958582, 214.950395, 214.97986, 214.827323, 214.802483], DEC [52.930525, 52.946655, 52.969334, 52.8194630, 52.837142]), that are bright and close to the galaxies. This method avoids smoothing due to stacking and implicitly accounts for local variations of the PSF.
Although all of these stars give very similar results, none of the stars is perfectly centered ($>$0.1 pixel shift in x and y). 

Finally, we inspect results from synthetic WebbPSF models. These fits are almost identical to those from single stars. Given the reproducability of WebbPSF, we use this for our reported measurements. The uncertainty in the PSF is taken into account in the error analysis, where we use both WebbPSF, single stars, and the stacked empirical PSF (see Section \ref{sec:results:sizesanduncertainties}). 

\begin{table*}[t!]

\small{
\begin{adjustwidth}{-2cm}{}
\begin{tabular}{llllllllllll}

RUBIES-ID & L23-ID & RA & DEC & $z_{\mathrm{spec}}$ & $\log$($M_{*,\mathrm{max}}$) &
$\log$($M_{*,\mathrm{med}}$) &
 $\log$($M_{*,\mathrm{min}}$) &
$\sigma$(H$\beta$) & $r_{\mathrm{e, maj}}$ & $b/a$ \\
& &  &  &  & [$\mathrm{M}_{\odot}$] &
[$\mathrm{M}_{\odot}$] &
 [$\mathrm{M}_{\odot}$] &
[km s$^{-1}$] & [pc] &  \\
\hline
49140 &  & 214.89225 & 52.87741 & 6.68 & 11.2$^{+0.15}_{-0.14}$ & 9.93$^{+0.22}_{-0.14}$ & 9.50$^{+0.08}_{-0.09}$ & 1402$^{+68}_{-79}$ & 123$^{+173}_{-45}$ &0.58 \\
55604 & 38904 & 214.98303 & 52.95600 & 6.98 & 11.1$^{+0.15}_{-0.16}$ & 9.79$^{+0.14}_{-0.14}$ & 8.99$^{+0.16}_{-0.14}$ & 1527$^{+106}_{-115}$ & 54$^{+53}_{-43}$ &0.78 \\
966323 & 14924 & 214.87615 & 52.88083 & 8.35 & 10.6$^{+0.13}_{-0.49}$ & 9.84$^{+0.20}_{-0.32}$ & 8.72$^{+0.21}_{-0.19}$ & 1369$^{+280}_{-315}$ & 86$^{+9}_{-42}$ &0.73 \\
\hline
\end{tabular}
\end{adjustwidth}
\caption{Spectroscopic and structural properties of three Balmer break galaxies. The first and second column indicate the galaxy ID given in \citet{Wang2024_balmer} and the corresponding ID number of L23. The spectroscopic redshift (column 4), stellar masses ($\mathrm{M}_{\odot}$) with $1\sigma$ errors (columns 4-6) and velocity dispersion (km s$^{-1}$) of the broad H$\beta$ line (column 7) are obtained from \citet{Wang2024_balmer}. The three different stellar masses are obtained by varying the AGN contribution relative to stellar contribution (see text).
The 8th and 9th column are the fitted  effective radii $r_{\rm e, maj}$ (pc) along the major axis and the minor-to-major axisratio ($b/a$), respectively. The galaxies are fit with a S\'ersic profile in filter F200W using WebbPSF. We perform simulations (see text) to obtain $1\sigma$ errors on the size measurements.
}
\label{tab:sizes}
}
\end{table*}

\subsection{Spectra}
The spectroscopic properties of the three LRDs\footnote{These sources have $m_{\rm F277W} - m_{\rm F444W}>1.5$, $m_{\rm F444W}<26$, similar to definitions of LRDs of \citet{Barro2024, PerezGonzalez2024A, Akins2024}. These sources also have broad Balmer emission lines, following the original criteria of \citet{Matthee2024}, even though that was not a selection criterion a priori.} with Balmer breaks shown in this Letter are obtained from \citet{Wang2024_balmer}. Briefly, the galaxies were selected from a spectroscopic follow-up program RUBIES \citep[JWST-GO-4233; PIs de Graaff
\& Brammer,][]{Graaff2024_rubies}, that uses the NIRSpec microshutter array (MSA). 
The RUBIES selection partially overlaps with the double break candidates of \citet{Labbe2023}.  The sources were targeted and observed in March 2024 for 48 min in the Prism/Clear mode and the G395M/F290LP mode. For more information on the reduction, we refer to \citet{Heintz2024}.

The modeling of the spectra is described in \citet{Wang2024_balmer}. 
Briefly, three different limiting cases are adopted to decompose the continuum into a stellar and AGN component. The first model maximizes the stellar contribution and minimizes the AGN contribution to the continuum, by setting it to zero.
The resulting stellar masses of all three galaxies are very high, $M_{\mathrm{*,max}}\sim10^{11}\mathrm{M}_\odot$. In the cumulative stellar mass density versus stellar mass plane, they lie close to the limit allowed by $\Lambda$CDM \citep[see both][]{BoylanKolchin2023,Wang2024_balmer}.
The second model assumes a maximal AGN contribution and therefore minimizes the contribution of stars. 
These minimal masses are $M_{\mathrm{*,min}}\sim10^{9}\mathrm{M}_\odot$.
As described in \citet{Wang2024_balmer}, this model leads to overmassive black holes relative to their host stellar masses, compared to the local relation.
They also adopt a third model, that lies in between the "maximum" and "minimum" cases, with stellar masses of $M_{\mathrm{*, med}}\sim10^{10}\mathrm{M}_\odot$. In this model, the AGN and stellar population both contribute equally to the red continuum at $\lambda_{\rm rest}\sim5000\rm\AA$, while the AGN dominates the red continuum beyond that wavelength.

While these three models are physically very different, they give equally good fits to the red continuum.
This is problematic for determining the stellar masses, which are not well constrained, with differences in stellar mass between the no-AGN and max AGN models of up to $\sim$2 dex.
Other caveats complicate the determination of stellar masses even further, such as the assumption of an IMF \citep[see e.g.][]{Dokkum2024} and the lack of information at near-infrared wavelengths from for example MIRI \citep{Akins2024, Wang2024_mirimasses}. These uncertainties are smaller than those associated with the choice of AGN contribution \citep[see e.g.][]{Conroy2009}, but we stress that there are significant systematic uncertainties even within each of the three \citet{Wang2024_balmer} models.
In this work,  we adopt the reported stellar masses from \citet{Wang2024_balmer}, derived from the spectra in rest-frame UV to optical and a \citet{Chabrier2003} IMF, but we stress the caveats involved in these stellar mass measurements.  Finally, we consider the three cases that vary the contribution of AGN as possible truths. The nine individual stellar
mass measurements are listed in Table 1. For more information on the fitting procedure and parameters in the modeling of the three scenarios, we refer to \citet{Wang2024_balmer}.

In Figure \ref{fig:spectra} we show the spectra in light grey and the best-fit model using the maximum stellar mass in dark blue. 
For illustration we also show the Balmer break at $\lambda_{\rm rest} = 3646\rm \AA$, the [OIII] line at $\lambda_{\rm rest} = 5007\rm \AA$ and H$\beta$ at $\lambda_{\rm rest} = 4861\rm \AA$ as dotted lines in black, red and blue, respectively.

\section{Structural Properties}

\subsection{Morphologies}
\label{sec:morphology}
In Figure \ref{fig:setup_segmentation} (top panels) we show the individual images ($1\arcsec \times 1\arcsec$) for the three Balmer break galaxies in F200W, as well as the NIRSpec slits
that were used.
We also show two RGB images, the first created from cutouts in F115W (blue), F150W (green), F200W (red), the second from F150W (blue), F277W (green), F444W (red). The images show that all three galaxies are extremely small, and appear as single, unresolved or barely resolved objects in the long wavelength bands. However, the morphologies of two of the galaxies
are more complex in the bluer bands.  ID-49140 and ID-55604 have a faint extension or small clump next to a compact primary component. These clumps are most prominent in F115W (see cutout images in Figure \ref{fig:spectra}).
The morphology of ID-966323 is more point-like in all bands.

\subsection{S\'ersic Profile Fitting}
\label{sec:methodology}
We fit the surface brightness profiles of the three Balmer break galaxies with  \citet{Sersic1968} profiles. 
As explained in \S\,\ref{sec:imaging} the fits are done in F200W, probing rest-frame
wavelengths of $\lambda_{\rm rest}\sim 200-250$\,nm. 
We use \textsc{galfit} \citep{Peng2002, Peng2010x}, and the
relevant
parameters are the central position ($x$, $y$), the effective radius (along the major axis) ($r_{\rm e}$), the S\'ersic index ($n$), the total integrated magnitude, projected minor-to-major axis ratio ($b/a$), and the position angle (PA). 

Initial fits showed that the S\'ersic index cannot be
determined robustly, given the small sizes of the
galaxies, but that it typically ranges between 1 - 3.5. We adopt a fixed $n=1.5$ in the fits. The uncertainty in $n$ translates into an uncertainty in $r_{\rm e}$, and this is propagated
in our error analysis (see Section \ref{sec:results:sizesanduncertainties}). The $r_{\rm e}$ is allowed to vary between 0.01 and 100 pixels and the total integrated magnitude between $-5$ and +5 magnitudes from the previously measured aperture magnitude.

We generate a mask with contaminating sources as follows. The background is estimated using sigma-clipped statistics with a filter size of 5 pixels and the background RMS ($\sigma$). 
After subtracting the background, the data are convolved with a 2D Gaussian kernel with a FWHM of 3 pixels. Using this convolved background-subtracted image, we adopt a source detection threshold of 1.5$\sigma$. The resulting map contains pixels that are not considered while fitting the galaxy models on the image with \textsc{galfit}.

\begin{figure}[t!]
    \centering
    \includegraphics[width=\linewidth, trim={2cm 0cm 0cm 0cm}, clip]{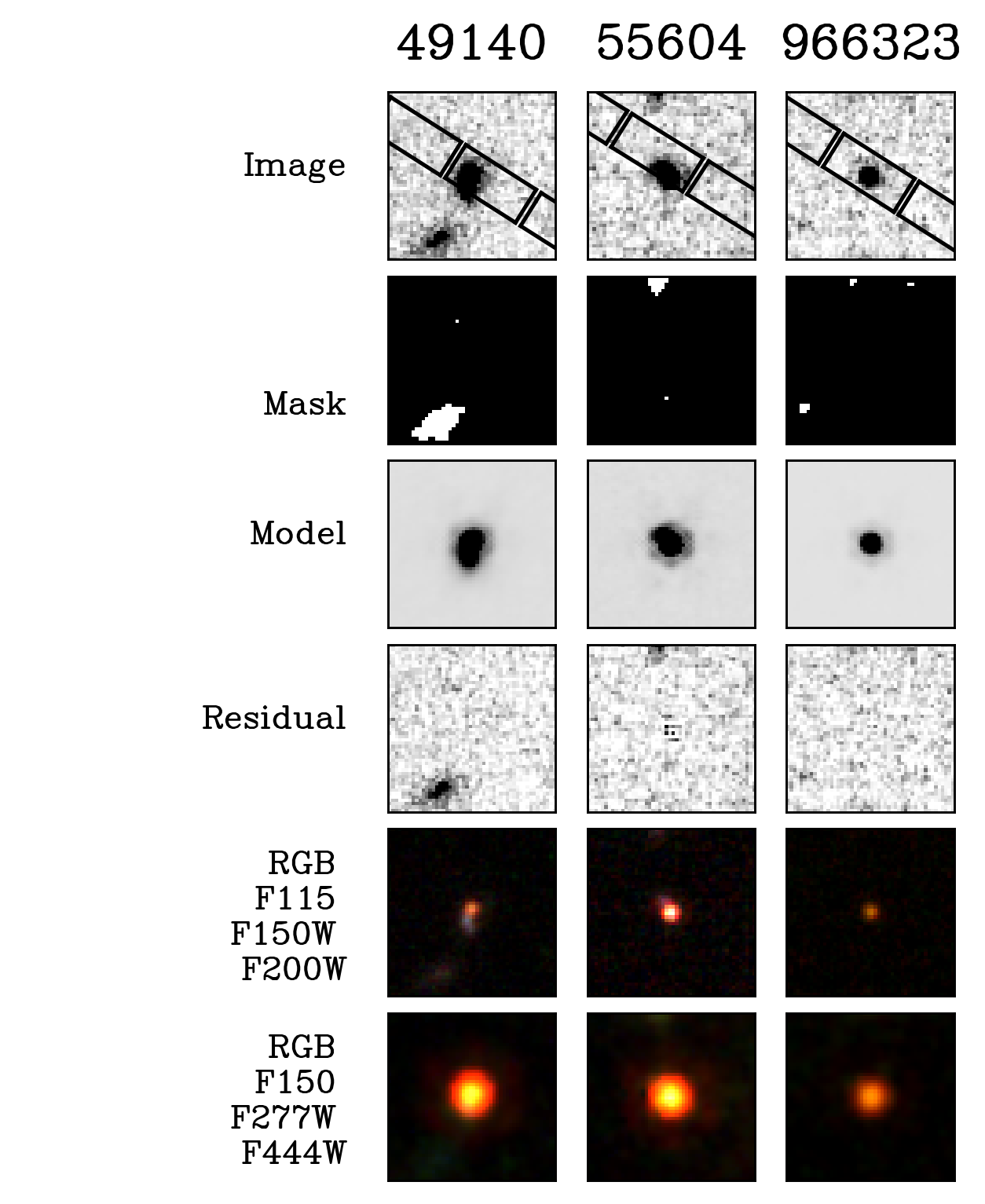}
    \caption{The top row shows images in the NIRCam F200W band of the three Balmer break galaxies. The second row shows the segementation maps, the third the best-fitting PSF convolved \textsc{galfit} models. The residuals, obtained by subtracting the best-fitting models from the images, are shown in the fourth row. The fifth row shows RGB images of the galaxies, with F115W as the blue band, F150W as the green band and F200W as the red band and the final row with F115W, F277W, F444W. All images are $1\arcsec \times 1\arcsec$.}
    \label{fig:setup_segmentation}
\end{figure}

For ID-55604 and ID-49140, we fit two S\'ersic components based on the images in F115W and F150W (see Section \ref{sec:morphology}). The nature of the relatively blue secondary, offset,
components is unclear. The primary component contains 70\,\% and
85\,\% of the total fluxes of ID-55604 and ID-49140,
respectively. As the primary components are redder than
the secondary components (particularly for ID-55604 -- see
Figure \ref{fig:setup_segmentation}), they contribute an even
larger fraction of the total masses.
Fitting the galaxies with single components leads to significant residuals, as expected, and sizes that are
twice as large as our default values. We include this
systematic uncertainty in the errorbars of these objects,
as discussed below.

\begin{figure*}[htp!]
    \centering
    \hspace*{3.2cm}\includegraphics[width=0.8\linewidth]{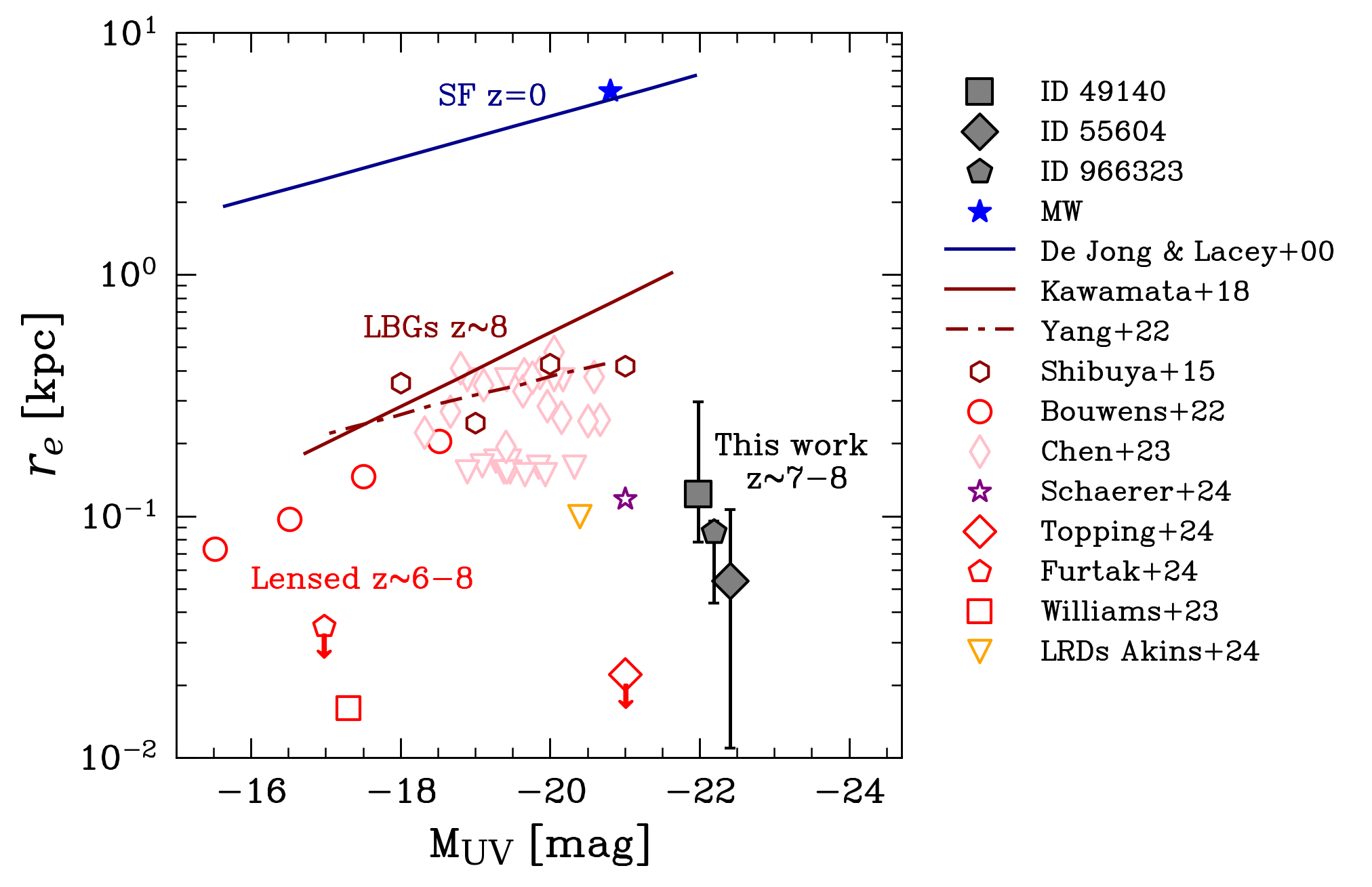}
    \caption{The size measured in F200W, plotted against the UV magnitude, measured from the spectrum at $\lambda_{\rm rest}$=150 nm. 
    We also show size-luminosity derived for local spiral galaxies from \citet{DeJong2000} for which we use a magnitude correction $\rm M_{\rm UV}$-$\rm M_{\rm I}=1.0$, and LBGs at $z\sim8$ \citep{Shibuya2015, Bouwens2022, Kawamata2018}, and $z\sim9-12$ \citep{Yang2022}. In addition we show star-forming complexes \citep{Chen2023} with $z_{\rm phot}\sim6-8$. Finally, we show extremely compact sources, recently detected through lensing with JWST, reported in \citet{Furtak2024} ($R_{\rm e}<35$ pc, $z=7.0$), \citet{Topping2024} ($R_{\rm e}<22$ pc, $z=6.1$) and \citet{Williams2023_science} ($R_{\rm e} = 16$ pc, $z=9.5$). Finally we show the upper limit of $\lesssim$100 pc reported for the brightest LRDs \citep{Akins2024} (see text).
    }
    \label{fig:sizemag}
\end{figure*}

\subsection{Sizes and Uncertainties}
\label{sec:results:sizesanduncertainties}

The masks, best-fitting models, and residuals from the fits
are shown in Figure 
\ref{fig:setup_segmentation}.
The fits are overall excellent. 
The effective radii are extremely small, $\sim$1.1 pixel or smaller. Using the spectroscopic redshifts
of the galaxies, we find physical effective radii of 123\,pc,
54\,pc, and 86\,pc for the three galaxies. 
To determine the uncertainties in these measurements we perform simulations to obtain upper and lower bounds for the sizes, taking into account the assumptions and unknowns in the modeling. 

For each galaxy
we consider a finely sampled array of possible sizes, from 0 to several pixels.
For each test size $r_{\rm e,\,test}$ we 
simulate $10^3$  S\'ersic models using \textsc{galfit}, with $r_{\rm e} =
r_{\rm e,\,test}$ and the same integrated magnitude of the best fit model, while randomly assigning $b/a$ from 0.1 to 1, PA from 0 to 90 degrees, and $n$ from 0.5 to 6. We convolve these S\'ersic models with either the stacked empirical PSF or WebbPSF. We then place these convolved models into the residual maps of the initial fits, so that differences between the galaxy profiles and S\'ersic profiles are included in the errors. 
We fit these synthetic images following the same methodology as was used to fit the galaxies. This includes keeping $n=1.5$ fixed, fitting with WebbPSF, and solving for $r_\mathrm{e}$, $b/a$, the integrated magnitude, PA, and the sky background. This results in an array of $10^3$ different sizes for which the underlying model had a true size of $r_{\rm e,\,test}$.  
Next, we ask whether the observed size $r_{\rm e,\,obs}$ is within the central 68\,\% of the distribution of $10^3$ simulated sizes. If so, then $r_{\rm e,\,test}$ is deemed an acceptable true size of the object, and it is included in the $1\sigma$ errorbar for that galaxy. 

We find that the uncertainties are asymmetric,
with smaller true sizes being more likely than larger true sizes. 
This can be traced to two effects. First, our choice of the WebbPSF for the default measurement: convolving a galaxy with an empirical PSF and then fitting it with WebbPSF leads to a slight ($\sim 0.2$\,pixel) overestimate of the size, and as half the simulations use empirical PSFs this leads to an overall bias in the simulated sizes. Second, there are more simulations with $n>1.5$ than with $n<1.5$, and fitting with fixed $n=1.5$ leads to a small (also $\sim 0.2$ pixels) bias. 

A final uncertainty that has to be taken into account is our choice of fitting
two galaxies (ID-55604 and ID-49140) with two components rather than one. We refit both galaxies with a single component to assess the importance of this choice. These fits lead to strong residuals, but they are stable.
The half-light radii of both galaxies increase by a factor of $\sim$2 when they are fit as a single component. To account for this, we add the difference  between the single-component and two-component fits in quadrature to the (positive) errorbars of these two galaxies. 
In Table \ref{tab:sizes} we report the measured sizes with their uncertainties. 

\subsection{Size-Luminosity Relation}
The central observational result of this Letter is the compactness of the three Balmer break galaxies at $z\sim7-8$. 
The sizes for ID-55604 and ID-966323 are lower than measured in \citet{Baggen2023} for identical galaxies due to the new reduction scheme (see Section \ref{sec:data}), but consistent within the errorbars. The average size of the three galaxies is 
$\langle r_{\rm e}\rangle \approx 90$ pc, smaller than the mean size of 150 pc in \citet{Baggen2023}. 

The sizes are similar to those of ultracompact dwarf galaxies  \citep[$r_{\mathrm{eff}}\leq$100 pc, e.g.][]{Zhang2015}, and smaller than
any other moderately luminous galaxy population observed at $0<z<5$. To illustrate the extreme nature of these objects,
we show the relation between effective radius (along the major axis) and UV magnitude ($\lambda_{\rm rest}\sim 1500$\,\AA) in Figure \ref{fig:sizemag}.  Besides the three
Balmer break galaxies
we show the relation for local spiral galaxies, obtained from \citet{DeJong2000}, for which we correct the measured $i$-band magnitudes to UV magnitude using $\rm M_{\rm UV}$-$\rm M_{\rm I} =1.0$, following \citet{Grazian2012}. 
We also show the size-luminosity relation for Lyman-break-galaxies (LBGs) detected prior to JWST at $z\sim8$ from \citet{Shibuya2015}, \citet{Bouwens2022}, \citet{Kawamata2018}, as well as recently detected galaxies with JWST at $z\sim9-12$ from \citet{Yang2022}.  
At fixed absolute magnitude, these galaxies are about
$100\times$ smaller than local spiral galaxies and about $10\times$ smaller than Lyman Break Galaxies (LBGs) at the same redshifts. Instead, their sizes are similar to those of lensed sources, reported in \citet{Bouwens2022}, which are $\sim 6$ magnitudes fainter. However, these sizes corroborate with some of the most recent results with JWST. We show the sizes (diamond) and upper limits (triangle) of star-forming complexes reported in \citet{Chen2023} with $z_{\rm phot}\sim6-8$. In addition, we show a bright compact galaxy at $z_{\rm spec}=9.4$ with $R_{\rm e} =118$ pc \citep{Schaerer2024_compact}. 
Extremely compact sources, recently detected through lensing with JWST, reported in \citet{Furtak2024} ($R_{\rm e}<35$ pc, $z=7.0$), \citet{Topping2024} ($R_{\rm e}<22$ pc, $z=6.1$), \citet{Williams2023_science} ($R_{\rm e} = 16$ pc, $z=9.5$) are shown as red scatter points. 
Finally, we show $r_{\rm e}\leq 100$ as a reference upper limit for brightest LRDs found by \citet{Akins2024} and we convert the reported flux at $\lambda_{\rm obs } = 1.15 \mu\rm m$ ($F = 15.76$ nJy) from the observed stacked spectrum to rest-frame UV magnitude, assuming a redshift of $z=6$.


\begin{figure*}[htp!]
    \centering
    \hspace*{3.0cm}
    \includegraphics[width=0.76\textwidth]{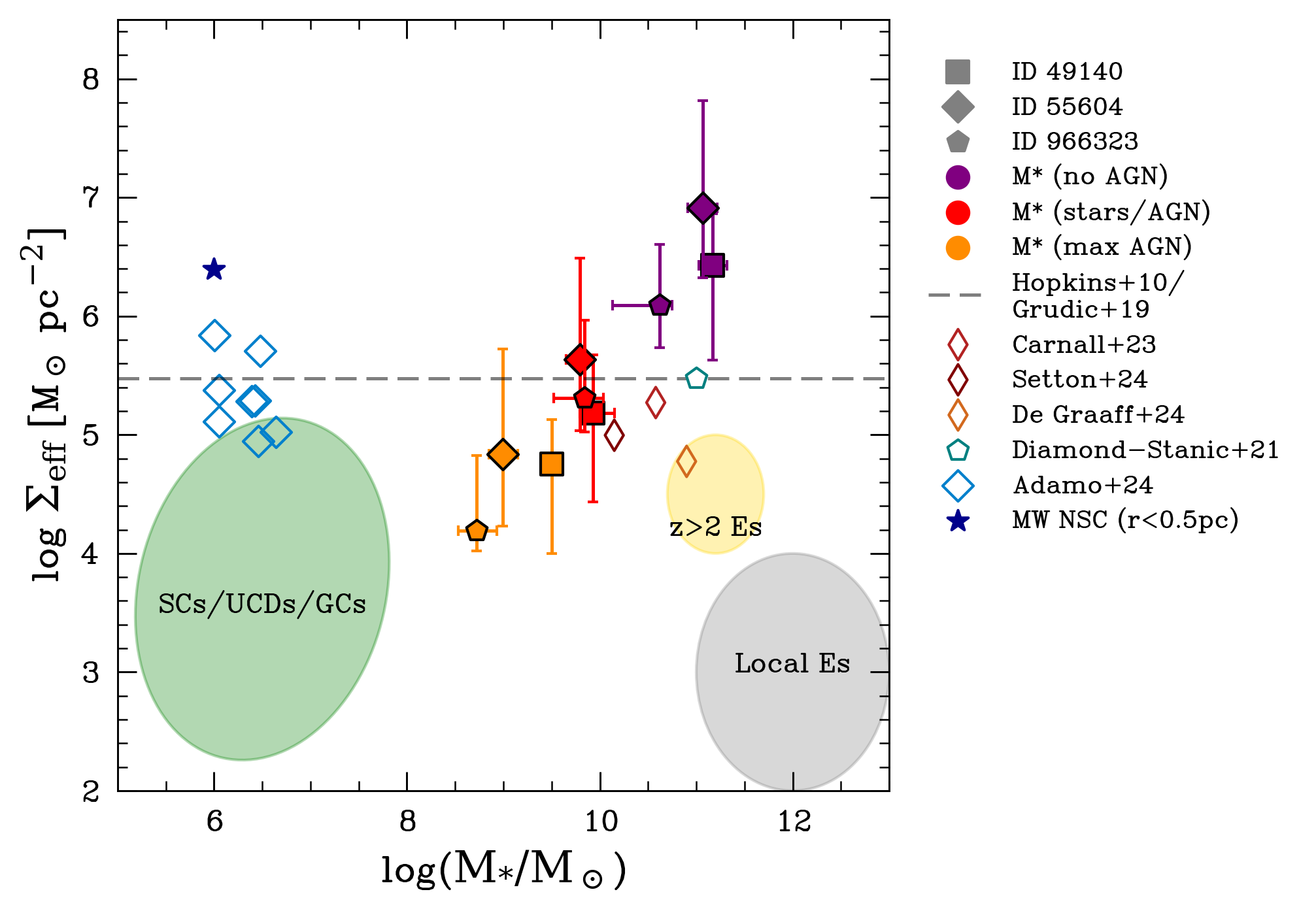}
    \caption{The surface density ($\Sigma_{\mathrm{eff}} = M_{*}(<R_{\mathrm{e}})/\pi R_{\mathrm{e}}^2$) plotted against stellar mass. 
    We show a theoretical limit $\Sigma_{\mathrm{eff}} \leqq 3\times 10^5 \mathrm{M}_{\odot} \mathrm{pc}^{-2}$, determined by \citet{Grudic2019} as the dashed black line.
    For illustration, we show regions in $M_*-\Sigma_{\mathrm{eff}}$ space for multiple stellar systems estimated from \citet{Hopkins2010_density}, the class of star clusters (SC), ultracompact dwarf galaxies (UCD) and globular clusters (GCs) (green),  local elliptical galaxies (grey) and compact $z>2$ elliptical galaxies (yellow). 
    We also show (see text) quiescent galaxies at $z\sim4-5$ \citep{Setton2024, Graaff2024, Carnall2023_nature}, compact starburst galaxies at $z=0.4-0.8$ \citep{Stanic2021}, Cosmic Gems star clusters at $z\sim10$ \citep{Adamo2024} and the MW nuclear star cluster. 
    The surface density measured for the three Balmer break galaxies are shown for the three sets of stellar masses $M_{*,\mathrm{min}}$, $M_{*,\mathrm{med}}$ and $M_{*,\mathrm{max}}$ in orange, red and purple, respectively. }    \label{fig:surfacedensity}
\end{figure*}

\section{Densities and Kinematics}

The small sizes, combined with the masses determined in \citet{Wang2024_balmer}, imply very high densities. Here we assess the average surface density of the galaxies (4.1), their 3D density profiles (4.2), and the expected kinematics (4.3).
In each subsection we consider all three mass measurements of \citet{Wang2024_balmer}, for minimal, medium, and maximal AGN contributions, and indicate those with different colors in the figures.

\subsection{Effective Surface Density}
\label{sec:stellardensities}
The average surface density within the effective radius can be calculated with
\begin{equation}
\label{eq:sigmaeff}
    \Sigma_{\mathrm{eff}} = M_{*}(<R_{\mathrm{e}})/\pi R_{\mathrm{e}}^2,
\end{equation}
where $M_{*}(<R_{\mathrm{e}})=M_*/2$ and $R_{\mathrm{e}}$ is the projected circularized half-stellar mass radius $R_{\rm e} = r_{\mathrm{e, maj}} \sqrt{b/a}$ (ignoring $M/L$ gradients).

In Figure \ref{fig:surfacedensity} we show the surface densities for the three galaxies, and for the three different stellar mass measurements of \citet{Wang2024_balmer}; $M_{*,\mathrm{min}}$, $M_{*,\mathrm{med}}$ and $M_{*,\mathrm{max}}$ in orange, red and purple, respectively. For context, we also show regions in $M_*-\Sigma_{\mathrm{eff}}$ space for various stellar systems, as estimated in \citet{Hopkins2010_density}. Nuclear star clusters, ultracompact dwarf galaxies, dSph nuclei and globular clusters are taken together and shown in green, local elliptical galaxies are shown in grey, and compact $z>2$ elliptical galaxies are shown in yellow. 
We also show recently discovered star clusters at $z\sim10$ \citep{Adamo2024} in the Cosmic Gems arc. 

In addition, we show the surface densities derived for three quiescent massive galaxies at $z=4-5$ from \citet{Carnall2023_nature}, \citet{Setton2024}, \citet{Graaff2024}. For these quiescent galaxies the effective radii are circularized using $b/a$=0.7, when not reported, adopting the value reported for the compact component in \citet{Setton2024}. Finally, we show the central surface densities for the compact starburst galaxies at $z=0.4-0.8$ \citep{Stanic2021}. 

The black dashed line shows an effective surface stellar mass density of $\Sigma_{\mathrm{eff}} \leqq 3\times 10^5 \mathrm{M}_{\odot} \mathrm{pc}^{-2}$. In the local Universe, very few stellar systems exceed 
this limit, going all the way
from star clusters and globular clusters to elliptical galaxies  \citep{Hopkins2010_density, Grudic2019}. 
It has been suggested that some universal mechanism controls the surface density: \citet{Hopkins2010_density} propose that 
feedback from massive stars in the form of winds and radiation fields is likely responsible for the observed $\Sigma_{\rm max}$. \citet{Grudic2019} propose an alternative model, relating the stellar feedback and star formation efficiency (SFE). 

The surface densities of the three galaxies discussed here are extremely high. 
They are above the empirical surface density
limit of \citet{Hopkins2010_density} by an order of magnitude in the no-AGN model, are on the limit for the mixed model, and are below the limit only for the maximal AGN model.
The densities are also higher by several
orders of magnitude 
than those of plausible descendants, elliptical
galaxies at $z=0$. 

The densities are most extreme for the
no-AGN model, and this could be taken as evidence against it.
However, we note that some nuclear star clusters reach $\sim10^6\mathrm{M}_{\odot} \mathrm{pc}^{-2}$ in their centers, above the empirical limit. An important example is the nuclear star cluster in the center of the Milky Way around SgrA$^*$, which reaches $2.5\times10^6\,\mathrm{M}_{\odot} \mathrm{pc}^{-2}$ within the central 0.5 pc \citep[see the review by][]{Neumayer2017}, indicated in Figure \ref{fig:surfacedensity} as a dark blue star. We calculated this surface density using a stellar mass within this region of $M_* \sim 10^6\mathrm{M}_{\odot}$ \citep{Schodel2009}.

\subsection{Stellar Mass Profiles}
\label{sec:discussion:stellarmassprofiles}
As first discussed in \citet{Bezanson2009}, 
extreme densities within the effective
radius do not necessarily correspond to
extreme densities on small physical scales.
The  effective radius evolves, and if galaxies grow inside-out, their
surface density within the effective radius goes down with time even if
their density within a fixed small physical radius remains constant.

\begin{figure}[t!]
    \centering
    \includegraphics[width=0.5\textwidth]{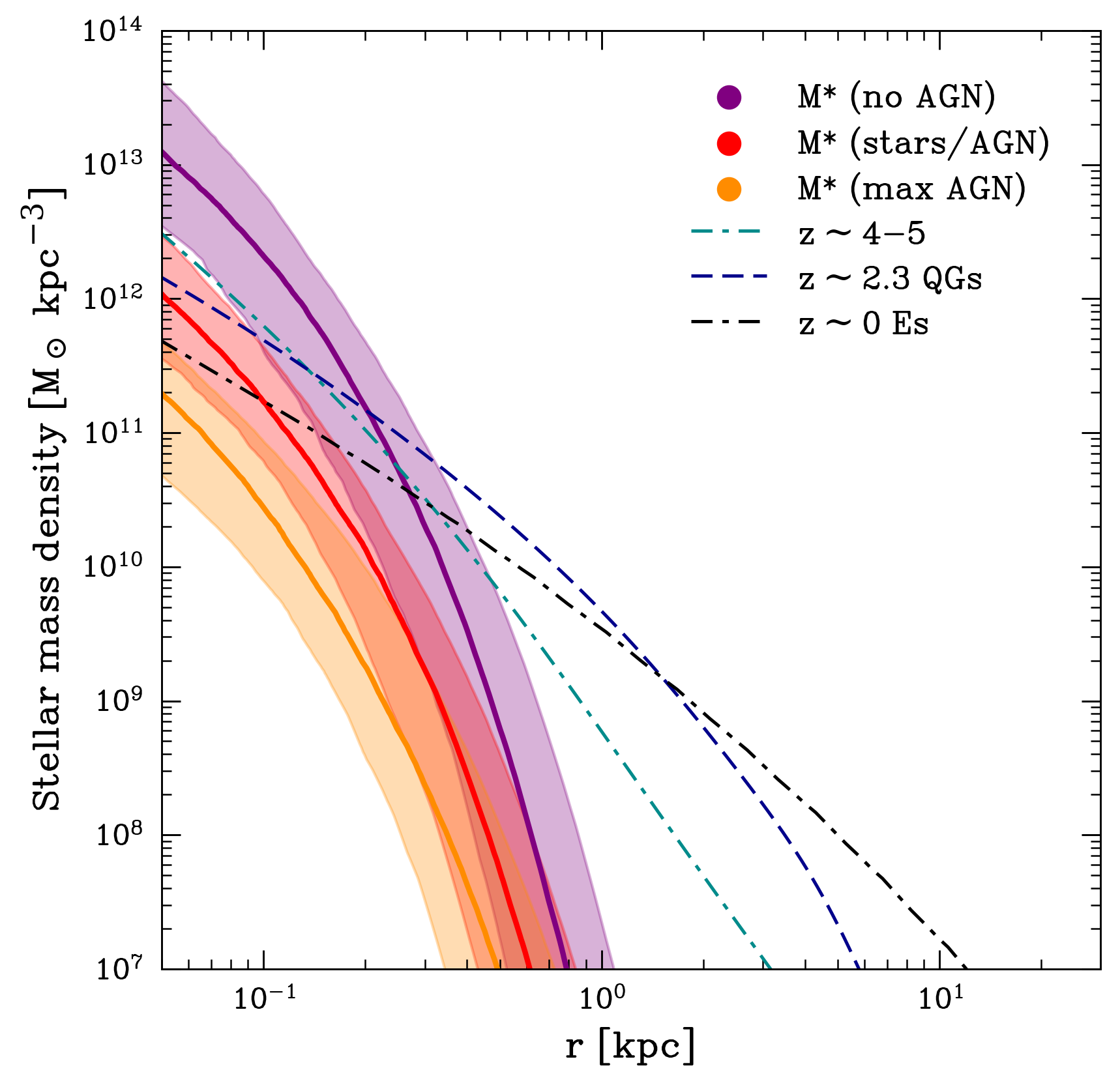}
    \caption{The median stellar mass profiles for the three sets of stellar masses (see Table \ref{tab:sizes}), $M_{*,\mathrm{min}}$, $M_{*,\mathrm{med}}$ and $M_{*,\mathrm{max}}$ in orange, red and purple, respectively. We also show the stellar mass profiles of massive quiescent galaxies at different cosmic times: the 
     mean stellar mass profile of three compact massive quiescent galaxies at $z\sim4-5$ \citep{Carnall2023_nature, Setton2024, Graaff2024}, $z\sim2.3$ compact elliptical galaxies \citep{Bezanson2009},
     and massive elliptical galaxies at $z=0$ \citep{Tal2009}.  
    }
    \label{fig:stellarmassdensity}
\end{figure}

Following \citet{Baggen2023} we show the 3D density profiles of the galaxies
in Figure \ref{fig:stellarmassdensity}. The profiles were determined from the radii and the three different sets of stellar masses, using the same methodology as in \citet{Baggen2023}. In short, we perform an Abel transform to the 2D best-fit S\'ersic profile (see Table \ref{tab:sizes}). After that, the luminosity profile is converted into a stellar mass profile by assuming the $M/L$ ratio does not change with radius. The profile is then scaled such that the total stellar masses are identical to those derived in
\citet{Wang2024_balmer}. The uncertainty bands reflect the uncertainties in both the radii and the masses. 
We also show the stellar mass profiles of massive quiescent galaxies at different cosmic times: $z\sim0$ elliptical galaxies \citep{Tal2009}, $z\sim2.3$ compact elliptical galaxies from \citet{Bezanson2009}, and the mean mass profile for three quiescent galaxies at $z\sim4-5$ \citep{Carnall2023_nature, Setton2024, Graaff2024}.

We find that the central stellar mass
densities are similar to those of plausible descendants for the
medium mass model (red points), and that they are lower for
the lowest masses (the maximal AGN model).
Strikingly, the central densities 
are extremely high for the no-AGN masses (purple): $\sim 10^{13}$\,M$_{\odot}$\,kpc$^{-3}$ in the inner tens of pc, comparable to the densest nuclear star clusters \citep{Pechetti2020}. They
are an order of magnitude above the density of plausible $z\sim0$ descendants, and also higher (by a factor of $\sim5$) than quiescent galaxies at $z=4-5$. If the no-AGN stellar masses are correct, it means that inside-out growth alone is
not sufficient to connect these galaxies
to their plausible descendants. 
The densities need to evolve downward, in order to be consistent with the central densities in the cores of quiescent galaxies at $z=0-5$. We return to
this in \S\,\ref{sec:discussion}.

\subsection{Kinematics}
\label{sec:broadlines}
\begin{figure*}[htp!]
    \centering
\includegraphics[width=0.6\textwidth]{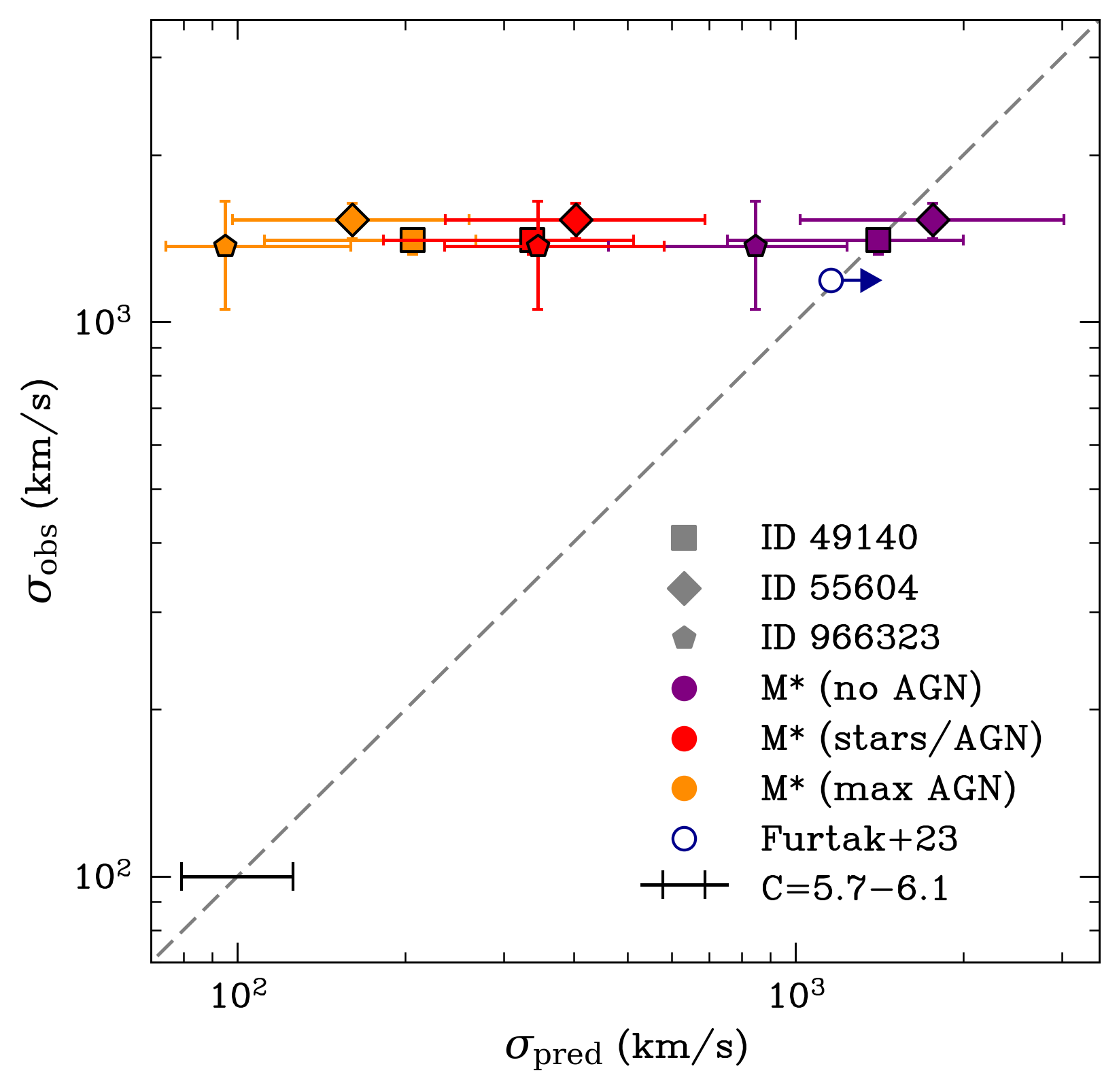}
     \hfill
\includegraphics[trim={22cm 2cm 20cm 5cm}, clip, width=0.39\textwidth]{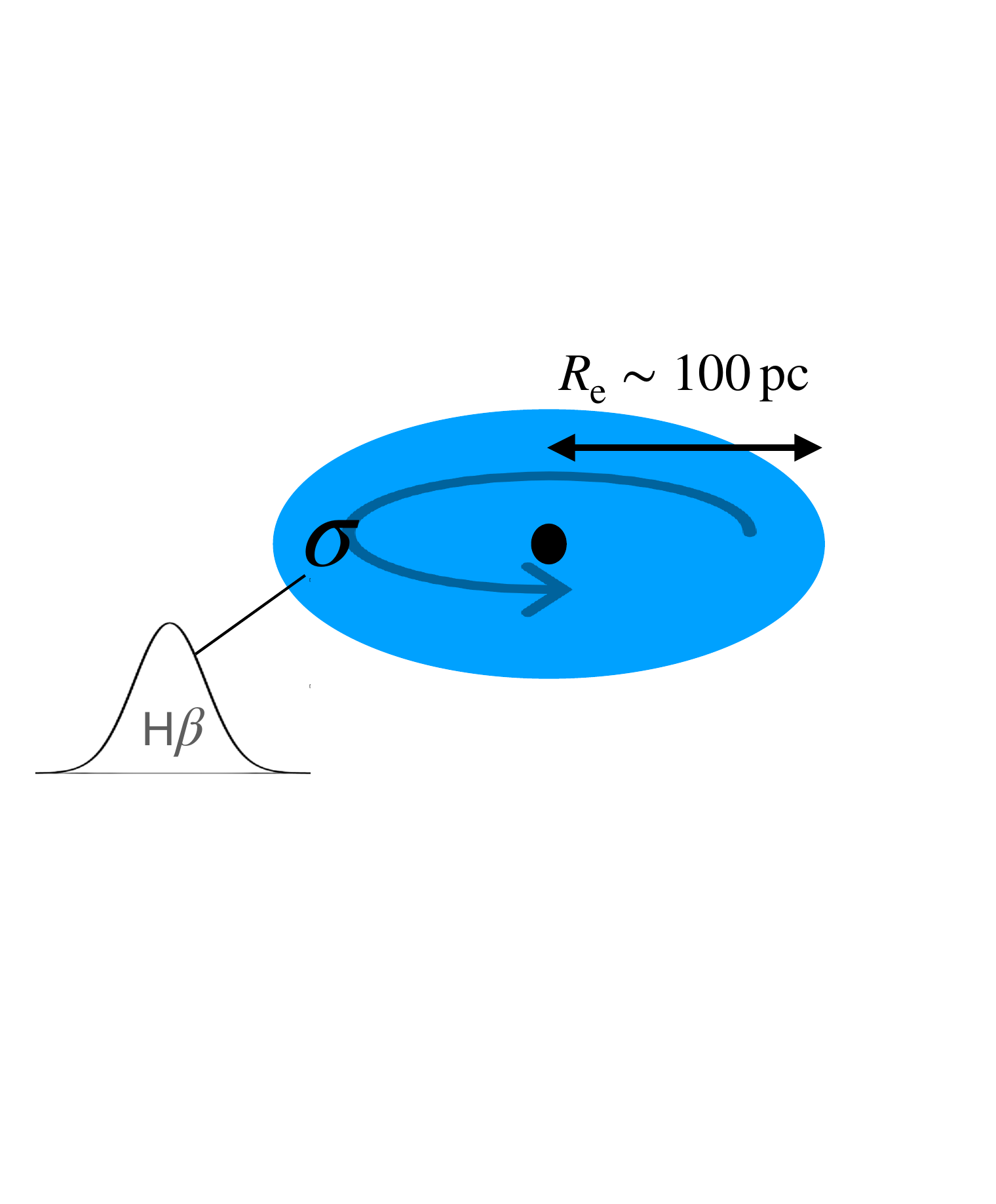}
    \caption{The observed velocity dispersion of the broad H$\beta$ lines plotted against the predicted velocity dispersion based on Eq.\ \ref{eq:veldisp} using the measured effective radius and 
    stellar masses (Table \ref{tab:sizes}). We show the three stellar mass models, $M_{*,\mathrm{min}}$, $M_{*,\mathrm{med}}$ and $M_{*,\mathrm{max}}$ in orange, red and purple, respectively. We predict high velocity dispersions, up to a few hundred km s$^{-1}$
    for the medium stellar mass scenario and $\sigma \gtrapprox$1000 km s$^{-1}$ for the maximum stellar mass scenario, similar to the observed velocities. The black errorbar in the lower left corner indicates how each measurement would shift horizontally when using different relations for Eq. \ref{eq:veldisp}, with $C=5.7$ (to the right) and $C=6.1$ (to the left). In the no-AGN model, the predicted velocity dispersions are very similar to the observed H$\beta$ line widths. Perhaps, 
    the broad Balmer emission lines
    in these galaxies could simply be the result of the small
    sizes and high masses of the galaxies. The right figure shows a schematic view of this scenario.
    }
    \label{fig:veldisp}
\end{figure*}

From a simple virial equilibrium argument, we can relate the dynamical mass ($M_{\rm dyn}$) of a galaxy to its velocity dispersion $\sigma$ at radius $r$:
\begin{equation} 
\label{eq:virialvelocity}
\sigma^2 \approx \frac{GM_{\rm dyn}(<r)}{kr}. 
\end{equation}
for which the virial coefficient $k$, depends on the galaxy structure and the shape of the velocity dispersion profile, 
and is typically assumed to be between $\sim3-5$ \citep[e.g.,][]{Cappellari2006,Franx2008,Taylor2010,vanderWel2006} 
If we make assumptions about the properties of the stars and gas in the system, following e.g.\ 
\citet{Dokkum2015}, we can also relate the velocity dispersion of the gas, to the stellar mass and circularized effective radius: 
\begin{equation}
\label{eq:veldisp}
 \log(\sigma_{\mathrm{pred}}) = 0.5 \left[\log(M_*) -\log(r_{\mathrm{e}}) - C\right],   
\end{equation}
where $\sigma_{\mathrm{pred}}$ in
$\mathrm{km \, s^{-1}}$, 
$M_*$ in units of $\mathrm{M}_\odot$, and $r_{\mathrm{e}}$ in kpc. 
The constant $C$ is uncertain, as it depends on the inclination and dynamics of the gas and stars, on their mass contributions, and on their relative
spatial distributions. 
Following \citet{Dokkum2015}  we adopt $C=5.9$, but 
to account for the uncertainties in the properties of the systems, we also consider the relations for $C=6.1$ \citep[as used in][]{Bezanson2009}, which corresponds to $1.25\times$ lower expected velocity dispersions for a given mass and size, and $C=5.7$ ($1.25\times$ higher dispersions).
For convenience, the full derivation along with all detailed assumptions is provided in Appendix \ref{app:veldisp}. 

In Figure \ref{fig:veldisp} we compare the
expected velocity dispersion calculated with Eq. \ref{eq:veldisp} to the measured widths of the H$\beta$ lines in the
three galaxies. The errorbars reflect the uncertainties in size and mass. The black errorbar in the lower left corner indicates how each measurement would shift horizontally when using $C=5.7$ (to the right) and $C=6.1$ (to the left). 
The expected dispersions are a strong function of
the choice of stellar mass, ranging from $\sim 150$\,km s$^{-1}$ for the minimal
masses to $\gtrsim 1000$\,km s$^{-1}$ for the maximal (no-AGN) mass.

Remarkably, the  expected dispersions are very similar to the
observed H$\beta$ line widths for the no-AGN model.
The implication is that the broad Balmer emission lines
in these galaxies could simply be the result of the small
sizes and high masses of the galaxies, rather than caused
by the broad line regions around supermassive black holes.
For reference, we also show the strongly-lensed red, compact
source reported in \citet{Furtak2023b, Furtak2024}, which
has a superficially similar spectrum as the three Balmer break objects in the present study. The upper limit to the effective radius is $R_{\rm e}\leq 35$\,pc, and the stellar mass for an SED dominated by stars is $\log(M_*/\mathrm{M}_{\odot}) = 10.57$\footnote{We note that this stellar mass is derived from photometry alone. Initial results of fitting the spectrum of this source with a stellar population show that the stellar mass is perhaps lower than this value \citep{Ma2024}.  } 
for this object. This gives an expected velocity dispersion (lower limit) similar to the reported value from the H$\beta$ line ($\sigma$\,=FWHM/2.35 =\,1190 \,km s$^{-1}$).

An important caveat is that the forbidden [O\,{\sc iii}]
lines in the three galaxies are narrow, of order $\sigma \sim 50-70$\,km s$^{-1}$.
In an AGN scenario this is explained by the high density of the
gas close to the black hole: the forbidden lines cannot
form there, because the collisional de-excitation rate exceeds the radiative de-excitation rate. Forbidden lines
arise from a narrow line region (NLR) at
larger distance. Interestingly, this same effect could
be at work in the no-AGN scenario. 
We derive the density of the gas as follows. If we assume the gas and the stars follow the same spatial distribution, $r_{\rm gas}\sim r_{\rm stars}\sim r_{\rm e}$, 
the ratio of the scale height to the scale length is $c/a \sim 0.5$, and $M_{\rm gas} \sim 0.5 M_{\rm stars}$, we obtain a gas density of $\rm N = 10^{6.0}-10^{6.9} \rm \,cm^{-3}$, which is larger than the critical density for the forbidden line \citep[$N_{\mathrm{cr, OIII, 5007\AA}}=10^{5.8} \mathrm{cm}^{-3}$,][]{Appenzeller1988}.
If the gas disk is thinner (i.e., $c/a<0.5$) the gas is even denser.\footnote{These results stem directly from the line
widths: the line width is a proxy for density, and for  $\sigma>1000$\,km s$^{-1}$ forbidden line formation is suppressed, irrespective of the spatial scale of the gas.}

\section{Discussion}
\label{sec:discussion}

The main observational result of this Letter is a confirmation
of the existence of extremely small ($r_{\rm e}\sim 100$\,pc),
luminous galaxies in the early Universe, at $z=6-8$.
Unlike other high
redshift galaxies, we know that the light at $\lambda_{\rm obs}
=2\,\mu$m is dominated by stars, thanks to the detection of Balmer
breaks in the spectra \citep{Wang2024_balmer}. 
Despite the accurate sizes and the Balmer break detections, the densities of the galaxies remain uncertain,
because AGN might contribute to the long wavelength flux
\citep[see][]{Wang2024_balmer}.

We show that the maximal mass (that is, no AGN) models produce objects that are an order of magnitude denser in their centers than any other known galaxies. This might be taken as evidence against
such models, except that they predict line widths that correspond remarkably well to the observed broad widths of the H$\beta$ lines.
If this is the correct interpretation for these galaxies, it
may apply to many other `little red dots' with broad emission
lines as well. 
It would provide a natural explanation for
the lack of X-ray detections, lack of variability, and relatively
faint MIRI fluxes of these galaxies (see the discussion and
references in the Introduction). 

It remains to be seen whether this scenario can explain
all aspects of these systems. 
First, the high luminosity in H$\beta$ ($L>10^{42}$  erg s$^{-1}$) may require an exotic mechanism for ionizing the gas. While AGNs are the most straightforward explanation, there are other possibilities. 
The extremely high density of the gas may
lead to collisional excitation or shocks being a significant contributor
\citep[e.g.][]{Davidson1985,Draine1993,Stasinska2001, StasisnkaSchaerer1999,Raga2015}. 
There is also some evidence for the presence of very massive stars at high redshift \citep{Upadhaya2024}, with higher ionizing power than typical stellar populations \citep[e.g.,][]{Schaerer2024}.
IMFs favoring the formation of massive stars
can also explain the population of bright galaxies at $z>10$ \citep{Harikane2023_imf,inayoshi2022,Menon2024, Trinca2024, Yung2024}. 
We note that alternative IMFs also change the stellar masses of the galaxies. In particular,
recently \citet{Dokkum2024} proposed a 'concordance IMF', with a steep low mass slope and a shallow high mass slope. The shape of this IMF is capable of reconciling both the high redshift bright, massive galaxies, and observations of the cores of the most massive galaxies in the local Universe.  This IMF produces slightly lower stellar masses ($\log(M_*)=-0.3$ to $-0.2$)
for the three galaxies than the \citet{Chabrier2003} IMF used by \citet{Wang2024_balmer}, but the change is much smaller than the effects of including AGNs in the modeling. 

Second, the [O\,{\sc iii}] lines should be spatially-extended
with respect to the Balmer lines. The spatial sampling
of the NIRSpec data ($0\farcs06-0\farcs1$) is, however, too coarse to confidently detect
this effect in the current data, but high resolution observations (with the NIRCam grism, or of a lensed system) should show this differential effect.
Third, the line profiles of the Balmer emission lines should
not be perfectly Gaussian but reflect rotation \citep[see, e.g.,][]{Dokkum2015}.
There is some evidence for asymmetries and perhaps two
peaks in the medium-resolution spectra of \citet{Wang2024_balmer}, but this could also be interpreted as absorption \citep[see e.g.][]{Wang2024a, Juodzbalis2024_rosetta}.
Finally, a stellar velocity dispersion measurement would be definitive. Such a case for an LRD has been reported very recently in \citet{Kokorev2024}, a little red dot with a massive ($\log(M_{*}/ \mathrm{M}_\odot) \sim 10.6$) quenched galaxy host which is also compact ($r_{\rm e}\leq300$ pc), for which a velocity dispersion of $\sigma\sim400$km s$^{-1}$ is predicted, identical to the observed broad stellar absorption lines. There is a broad H$\beta$ absorption component, but it is too weak to robustly determine the width of the line. 
A further test for these Balmer break galaxies is that for $\sigma>1000$\,km s$^{-1}$ the blended absorption
lines in the Balmer break
region take on a smoother shape than for ``normal'' velocity
dispersions, and this should be detectable in high S/N spectra. 

\begin{figure*}[htp!]
    \centering
    \includegraphics[width=0.6\linewidth]{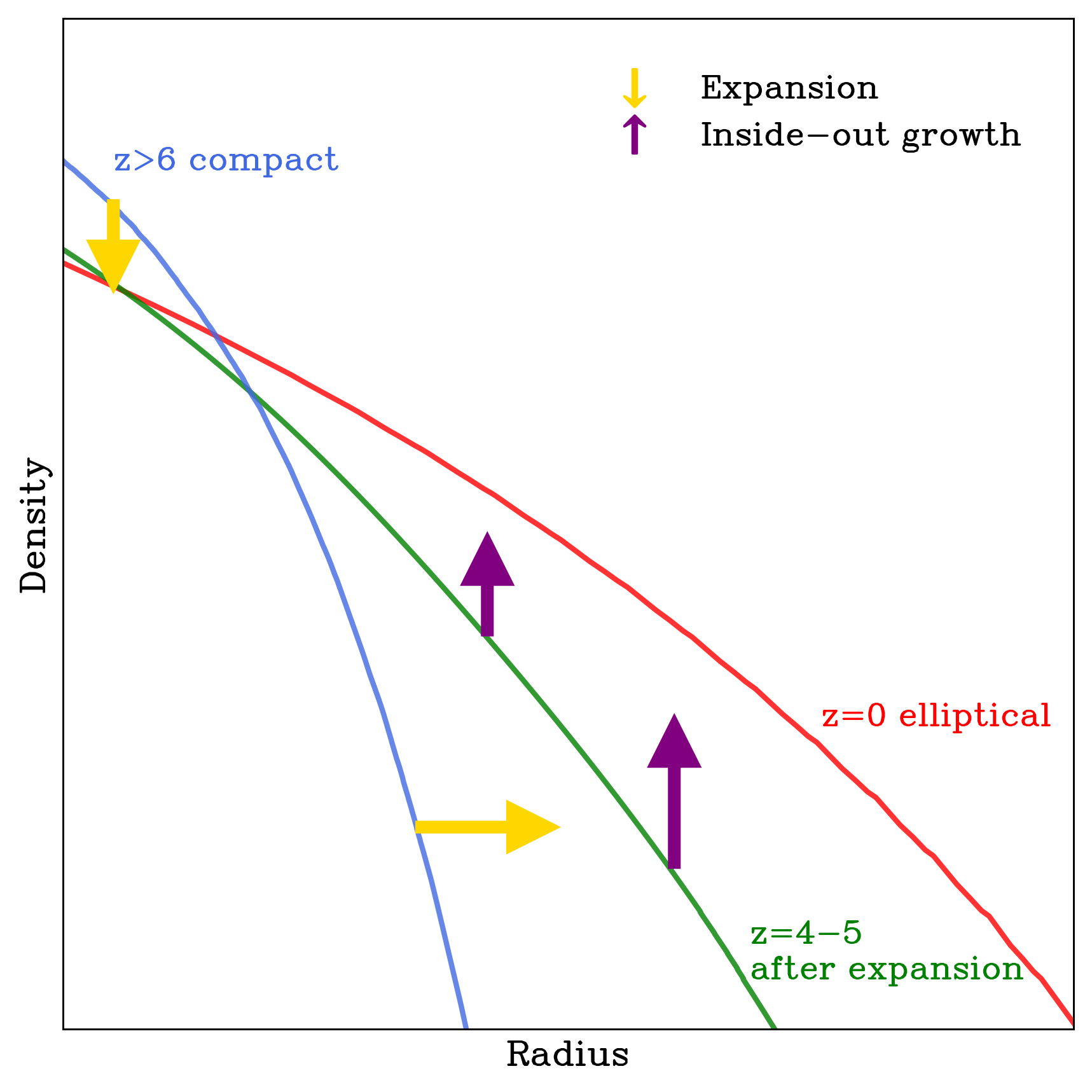}
    \caption{A possible pathway of the massive compact galaxies at $z>6$ in the no-AGN scenario. First, a phase of expansion as a result of mass loss \citep[triggered by feedback e.g.][]{Fan2008} or interaction of the SMBH with the stars in the central region 
    \citep[e.g.][]{Begelman1980, Ebisuzaki1991,Hills1983,Quinlan1996}
     is required to lower the densities and velocity dispersions to those of massive compact quiescent galaxies at $z=4-5$. After that the galaxies grow mainly in an inside-out fashion, due to minor mergers \citep[e.g.][]{Naab2009,Bezanson2009, Dokkum2010}.  }
    \label{fig:cartoon_densities}
\end{figure*}

If this interpretation were confirmed,
key questions are what might have caused such extremely small and dense galaxies to form, and how they evolve into ``normal''
systems with larger sizes and lower central densities. 
Perhaps the conditions in these systems are 
such that feedback-free starbursts (FFB) can occur \citep{Dekel2023}, 
with the lack of feedback leading to extreme star formation efficiencies and thereby extreme stellar masses. FFB galaxies are indeed expected to be compact \citep{Li2024}. 
It was also recently suggested that the compactness may simply be driven by the steepness of dark matter profiles at $z=10-14$ \citep{BoylanKolchin2024}. Their compactness could also be the result of the low angular momentum of halos during their initial collapse, as proposed by \citet{EisensteinLoeb1995, Loeb2024}.
Some general hydrodynamical simulations also produce small sizes for the most massive galaxies, such as $z\sim6$ galaxies in TNG50 \citep{Costantin2023} and $z\sim 7$ galaxies in the BLUETIDES simulation  \citep{Marshall2022}.    In addition, \citet{Roper2023} find 
 that bulge formation begins by efficient cooling and high star formation rates in the cores down to effective radii of $r_{\rm e}\sim100$\,pc in the FLARES simulation. 
However, others find discrepancies between the observed and simulated sizes of high-redshift galaxies \citep[e.g.\ THESAN,][]{Shen2024}. It remains an open question why compactness seems to be such a generic feature of early bright star-forming galaxies
\citep[$z\sim7-9$, $r_{\rm e}\sim100-300$\,pc, e.g.][]{Baggen2023, Akins2023, Langeroodi2023, Baker2024, Schaerer2024_compact, Ono2022} and massive quiescent galaxies out to $z\sim5$   \citep[$r_{\rm e}\sim200-500$\,pc, e.g.][]{Ji2024, Wright2024, Setton2024, Carnall2023_nature, Graaff2024, Kokorev2024}.

Whatever the formation mechanism, in the no-AGN scenario, dramatic evolution is required to bring the stellar mass densities in line with those of plausible descendants. One possible mechanism is adiabatic expansion due
to feedback-driven mass loss, from AGNs or stellar winds \citep{Fan2008}. Feedback may be suppressed at early times \citep{Dekel2023} and then become very effective once it is ``turned on''.
Another mechanism is scouring by a binary SMBH \citep[e.g.][]{Begelman1980, Ebisuzaki1991,Hills1983,Quinlan1996}, after a merger. The ejection of stars from the core lowers the density, increases the size, and likely also reduces rotational support \citep[e.g.,][]{Rantala2024}. 
While these models have been proposed to explain the size evolution of lower redshift ($z\sim2$) compact red nuggets, as well as the formation of cores in present-day ellipticals, they work on the right spatial scales to be relevant here.
An attractive feature of such scenarios is that they act quickly, lower the central density and increase the effective radius, and
keep the stellar mass roughly constant,
bringing the galaxies close to observed quiescent
galaxies at $z=4-5$ \citep{Setton2024, Carnall2023_nature, Graaff2024}. 
After this expansion phase the galaxies slowly grow
inside-out, largely due to minor mergers \citep[see, e.g.,][]{Naab2009,Dokkum2010}. These pathways are illustrated in
Figure \ref{fig:cartoon_densities}.

It also remains to be seen whether these ultra compact galaxies
are stable against gravitational collapse; it may that parts
of the galaxies collapse into supermassive black holes while
the rest expands (\citealt{LyndenbellWood1968}, and see also \citealt{Dekel2024}, \citealt{Kroupa2020}).

Finally, we stress that the standard interpretation of the
broad Balmer emission lines, namely broad line regions
around supermassive black holes, remains viable --- and perhaps
more likely. 
Future observations may show whether the
continuum is dominated by AGN light or stellar light, for instance through probing the red continuum beyond $\lambda_{\rm rest}=1.6\,\mu$m with ALMA/MIRI
\citep[see, e.g.,][]{Akins2024, Labbe2023b,Iani2024,  PerezGonzalez2024A,Williams2024}. 
The no-AGN solution that we explore here is self-consistent and
intriguing, but it requires further evidence.

\begin{acknowledgements}
\section*{Acknowledgements}
We would like to thank Pablo P\'erez-Gonz\'alez, Aayush Saxena and Avishai Dekel for useful discussions. 
We are also thankful for the observations made with the NASA/ESA/CSA James Webb Space Telescope and the 
 Cosmic Evolution Early Release Science (CEERS) program \citep[PI: Finkelstein; PID: 1345][]{Finkelstein2022, Finkelstein2023} \citep[see data DOI:][]{doiCEERSmosaic}. 
The CEERS data are publicly available in the Mikulski Archive for Space Telescopes (MAST) archive at the Space Telescope Science Institute 
(DOI: \dataset[https://doi.org/10.17909/z7p0-8481]{https://doi.org/10.17909/z7p0-8481}). STScI is operated by the Association of Universities for Research in Astronomy, Inc., under NASA contract NAS5–26555.  The \textsc{grizli} pipeline was used to reduce the data, which are available through the Dawn JWST Archive (DJA). DJA is an initiative of the Cosmic Dawn Center, which is funded by the Danish National Research Foundation under grant No.\ 140.
Finally, we thank the anonymous referee for their useful comments on this paper, which significantly enhanced the quality of our work. 
\end{acknowledgements}

\appendix
\section{Expected velocity dispersion}

\label{app:veldisp}
In this appendix, we show the derivation of Eq. \ref{eq:veldisp}, following \citet{Dokkum2015} and references therein.

The observed velocity dispersion of the gas, which is the second moment of the velocity distribution of the gas along the line of sight, for an unresolved rotating disk is given by: 
\begin{equation}
    \sigma^2_{\rm gas} = \alpha^2 V_{\rm rot}^2 \sin^2{i} + \sigma_{\rm random}^2,
\end{equation}
where $i$ is the inclination (such that face-on systems have $i=0$ and edge-on disks have $i=90$), 
$\alpha$ is typically $\sim0.6-1$ \citep{Franx1993, Rix1996, Weiner2006}. The random motion term contains the dispersions within the gas clouds in the interstellar medium and inclination dependent non-gravitational motions such as winds:  
$\sigma_{\rm random}^2 = \sigma^2_{\rm ism} + w^2(i) \sigma^2_{\rm wind}$. 
The rotation velocity ($V_{\rm rot}$) of the gas at a radius $r_{\rm gas}$ is related via the virial theorem to the dynamical mass 
enclosed in $r_{\rm gas}$: 
\begin{equation}
    V_{\rm rot, gas}^2 = \frac{G M_{\rm dyn}(<r_{\rm gas})}{ k \, r_{\rm gas}},
\end{equation}
where $k$ is the (typically radius-dependent) virial coefficient. For a perfectly spherical case, $k=1$, but for a thin disk, a value of $k=1.8$ (corresponding to $n=1$) is typically used \citep[see e.g.][]{Rowland2024}. 
We assume that the stars in the compact galaxies in this work formed simultaneously with the gas, such that they have the same distribution and
kinematics. Therefore, we assume that the size of the gas disk is equal to that of the stars, $r_{\mathrm{gas}}  =r_{\mathrm{stars}} =r_{\mathrm{e}}$. In principle, this might not be true at all \citep[see discussion in][]{Dokkum2015}; the gas could either be more extended, or more compact. 
If we also ignore the random motions of the gas, we obtain:


\begin{equation}
    \sigma^2_{\rm gas} = \alpha^2 \sin^2{(i)} \frac{G M_{\rm dyn}(<r_{\rm e})}{k r_{\rm e}}
\end{equation}

Assuming that the dynamical mass consists of gas and stars, neglecting dark matter in the center, 
we have $M_{\rm dyn}(<r_{\rm e}) = M_{\rm gas}(<r_{\rm e}) + M_{\rm *}(<r_{\rm e})$.
If we then assume the mass fraction in gas and stars to be equal ($f_{\rm gas}=0.5$), we have $M_{\rm dyn}(<r_{\rm e}) = 2 M_{\rm *}(<r_{\rm e})$, where the half light radius contains half the stellar mass so $M_{\rm dyn}(<r_{\rm e}) = M_*$. 
Plugging in numbers for $\alpha\sim0.8$, $k=1$, $i=45^{\circ}$ and $G = 4.3 \times 10^{-6}$ kpc ($\mathrm{M}_\odot)^{-1}$ $(\rm km \rm \, s^{-1})^2$,
we have the expected velocity dispersion of the gas, as a function of stellar mass and effective radius: 
\begin{equation}
     \log \sigma_{\rm pred} = 0.5\left[ \log(M_*) - \log(r_{\rm e}) -5.9\right].
\end{equation}

It is clear that the constant $C$ can vary, depending on the assumed properties of the system. For example, if we assume $k=1.8$ instead of $k=1$, we get $C=6.1$ (x1.25 lower velocity dispersion), while an inclination of $i=60^{\circ}$ would lead to $C=5.7$ (x1.25 higher velocity dispersion).

\bibliography{paper}{}

\begin{thebibliography}{}
\expandafter\ifx\csname natexlab\endcsname\relax\def\natexlab#1{#1}\fi
\providecommand{\url}[1]{\href{#1}{#1}}
\providecommand{\dodoi}[1]{doi:~\href{http://doi.org/#1}{\nolinkurl{#1}}}
\providecommand{\doeprint}[1]{\href{http://ascl.net/#1}{\nolinkurl{http://ascl.net/#1}}}
\providecommand{\doarXiv}[1]{\href{https://arxiv.org/abs/#1}{\nolinkurl{https://arxiv.org/abs/#1}}}

\bibitem[{{Adamo} {et~al.}(2024){Adamo}, {Bradley}, {Vanzella}, {Claeyssens}, {Welch}, {Diego}, {Mahler}, {Oguri}, {Sharon}, {Abdurro'uf}, {Hsiao}, {Xu}, {Messa}, {Lassen}, {Zackrisson}, {Brammer}, {Coe}, {Kokorev}, {Ricotti}, {Zitrin}, {Fujimoto}, {Inoue}, {Resseguier}, {Rigby}, {Jim{\'e}nez-Teja}, {Windhorst}, {Hashimoto}, \& {Tamura}}]{Adamo2024}
{Adamo}, A., {Bradley}, L.~D., {Vanzella}, E., {et~al.} 2024, \nat, 632, 513, \dodoi{10.1038/s41586-024-07703-7}

\bibitem[{{Akins} {et~al.}(2023){Akins}, {Casey}, {Allen}, {Bagley}, {Dickinson}, {Finkelstein}, {Franco}, {Harish}, {Arrabal Haro}, {Ilbert}, {Kartaltepe}, {Koekemoer}, {Liu}, {Long}, {McCracken}, {Paquereau}, {Papovich}, {Pirzkal}, {Rhodes}, {Robertson}, {Shuntov}, {Toft}, {Yang}, {Barro}, {Bisigello}, {Buat}, {Champagne}, {Cooper}, {Costantin}, {de La Vega}, {Drakos}, {Faisst}, {Fontana}, {Fujimoto}, {Gillman}, {G{\'o}mez-Guijarro}, {Gozaliasl}, {Hathi}, {Hayward}, {Hirschmann}, {Holwerda}, {Jin}, {Kocevski}, {Kokorev}, {Lambrides}, {Lucas}, {Magdis}, {Magnelli}, {McKinney}, {Mobasher}, {P{\'e}rez-Gonz{\'a}lez}, {Rich}, {Seill{\'e}}, {Talia}, {Urry}, {Valentino}, {Whitaker}, {Yung}, {Zavala}, {Cosmos-Web Team}, \& {Ceers Team}}]{Akins2023}
{Akins}, H.~B., {Casey}, C.~M., {Allen}, N., {et~al.} 2023, \apj, 956, 61, \dodoi{10.3847/1538-4357/acef21}

\bibitem[{{Akins} {et~al.}(2024){Akins}, {Casey}, {Lambrides}, {Allen}, {Andika}, {Brinch}, {Champagne}, {Cooper}, {Ding}, {Drakos}, {Faisst}, {Finkelstein}, {Franco}, {Fujimoto}, {Gentile}, {Gillman}, {Gozaliasl}, {Harish}, {Hayward}, {Hirschmann}, {Ilbert}, {Kartaltepe}, {Kocevski}, {Koekemoer}, {Kokorev}, {Liu}, {Long}, {McCracken}, {McKinney}, {Onoue}, {Paquereau}, {Renzini}, {Rhodes}, {Robertson}, {Shuntov}, {Silverman}, {Tanaka}, {Toft}, {Trakhtenbrot}, {Valentino}, \& {Zavala}}]{Akins2024}
{Akins}, H.~B., {Casey}, C.~M., {Lambrides}, E., {et~al.} 2024, arXiv e-prints, arXiv:2406.10341, \dodoi{10.48550/arXiv.2406.10341}

\bibitem[{{Ananna} {et~al.}(2024){Ananna}, {Bogd{\'a}n}, {Kov{\'a}cs}, {Natarajan}, \& {Hickox}}]{Ananna2024}
{Ananna}, T.~T., {Bogd{\'a}n}, {\'A}., {Kov{\'a}cs}, O.~E., {Natarajan}, P., \& {Hickox}, R.~C. 2024, \apjl, 969, L18, \dodoi{10.3847/2041-8213/ad5669}

\bibitem[{{Appenzeller} \& {Oestreicher}(1988)}]{Appenzeller1988}
{Appenzeller}, I., \& {Oestreicher}, R. 1988, \aj, 95, 45, \dodoi{10.1086/114611}

\bibitem[{{Baggen} {et~al.}(2023){Baggen}, {van Dokkum}, {Labb{\'e}}, {Brammer}, {Miller}, {Bezanson}, {Leja}, {Wang}, {Whitaker}, {Suess}, \& {Nelson}}]{Baggen2023}
{Baggen}, J. F.~W., {van Dokkum}, P., {Labb{\'e}}, I., {et~al.} 2023, \apjl, 955, L12, \dodoi{10.3847/2041-8213/acf5ef}

\bibitem[{{Baker} {et~al.}(2024){Baker}, {Tacchella}, {Johnson}, {Nelson}, {Suess}, {D'Eugenio}, {Curti}, {de Graaff}, {Ji}, {Maiolino}, {Robertson}, {Scholtz}, {Alberts}, {Arribas}, {Boyett}, {Bunker}, {Carniani}, {Charlot}, {Chen}, {Chevallard}, {Curtis-Lake}, {Danhaive}, {DeCoursey}, {Egami}, {Eisenstein}, {Endsley}, {Hausen}, {Helton}, {Kumari}, {Looser}, {Maseda}, {Pusk{\'a}s}, {Rieke}, {Sandles}, {Sun}, {{\"U}bler}, {Williams}, {Willmer}, \& {Witstok}}]{Baker2024}
{Baker}, W.~M., {Tacchella}, S., {Johnson}, B.~D., {et~al.} 2024, Nature Astronomy, \dodoi{10.1038/s41550-024-02384-8}

\bibitem[{{Balogh} {et~al.}(1999){Balogh}, {Morris}, {Yee}, {Carlberg}, \& {Ellingson}}]{Balogh1999}
{Balogh}, M.~L., {Morris}, S.~L., {Yee}, H.~K.~C., {Carlberg}, R.~G., \& {Ellingson}, E. 1999, \apj, 527, 54, \dodoi{10.1086/308056}

\bibitem[{{Barro} {et~al.}(2024){Barro}, {P{\'e}rez-Gonz{\'a}lez}, {Kocevski}, {McGrath}, {Trump}, {Simons}, {Somerville}, {Yung}, {Arrabal Haro}, {Akins}, {Bagley}, {Cleri}, {Costantin}, {Davis}, {Dickinson}, {Finkelstein}, {Giavalisco}, {G{\'o}mez-Guijarro}, {Hathi}, {Hirschmann}, {Holwerda}, {Huertas-Company}, {Kartaltepe}, {Koekemoer}, {Lucas}, {Papovich}, {Pirzkal}, {Seill{\'e}}, {Tacchella}, {Wuyts}, {Wilkins}, {de la Vega}, {Yang}, \& {Zavala}}]{Barro2024}
{Barro}, G., {P{\'e}rez-Gonz{\'a}lez}, P.~G., {Kocevski}, D.~D., {et~al.} 2024, \apj, 963, 128, \dodoi{10.3847/1538-4357/ad167e}

\bibitem[{{Begelman} {et~al.}(1980){Begelman}, {Blandford}, \& {Rees}}]{Begelman1980}
{Begelman}, M.~C., {Blandford}, R.~D., \& {Rees}, M.~J. 1980, \nat, 287, 307, \dodoi{10.1038/287307a0}

\bibitem[{{Bezanson} {et~al.}(2009){Bezanson}, {van Dokkum}, {Tal}, {Marchesini}, {Kriek}, {Franx}, \& {Coppi}}]{Bezanson2009}
{Bezanson}, R., {van Dokkum}, P.~G., {Tal}, T., {et~al.} 2009, \apj, 697, 1290, \dodoi{10.1088/0004-637X/697/2/1290}

\bibitem[{{Bogd{\'a}n} {et~al.}(2024){Bogd{\'a}n}, {Goulding}, {Natarajan}, {Kov{\'a}cs}, {Tremblay}, {Chadayammuri}, {Volonteri}, {Kraft}, {Forman}, {Jones}, {Churazov}, \& {Zhuravleva}}]{Bogdan2024}
{Bogd{\'a}n}, {\'A}., {Goulding}, A.~D., {Natarajan}, P., {et~al.} 2024, Nature Astronomy, 8, 126, \dodoi{10.1038/s41550-023-02111-9}

\bibitem[{{Bouwens} {et~al.}(2022){Bouwens}, {Illingworth}, {van Dokkum}, {Oesch}, {Stefanon}, \& {Ribeiro}}]{Bouwens2022}
{Bouwens}, R.~J., {Illingworth}, G.~D., {van Dokkum}, P.~G., {et~al.} 2022, \apj, 927, 81, \dodoi{10.3847/1538-4357/ac4791}

\bibitem[{{Boylan-Kolchin}(2023)}]{BoylanKolchin2023}
{Boylan-Kolchin}, M. 2023, Nature Astronomy, 7, 731, \dodoi{10.1038/s41550-023-01937-7}

\bibitem[{{Boylan-Kolchin}(2024)}]{BoylanKolchin2024}
---. 2024, arXiv e-prints, arXiv:2407.10900.
\newblock \doarXiv{2407.10900}

\bibitem[{{Brammer}(2023)}]{gabe_brammer_grizli}
{Brammer}, G. 2023, grizli, 1.5.2,  Zenodo, \dodoi{10.5281/ZENODO.1146904}

\bibitem[{{Bruzual A.}(1983)}]{Bruzual1983}
{Bruzual A.}, G. 1983, \apj, 273, 105, \dodoi{10.1086/161352}

\bibitem[{{Cappellari} {et~al.}(2006){Cappellari}, {Bacon}, {Bureau}, {Damen}, {Davies}, {de Zeeuw}, {Emsellem}, {Falc{\'o}n-Barroso}, {Krajnovi{\'c}}, {Kuntschner}, {McDermid}, {Peletier}, {Sarzi}, {van den Bosch}, \& {van de Ven}}]{Cappellari2006}
{Cappellari}, M., {Bacon}, R., {Bureau}, M., {et~al.} 2006, \mnras, 366, 1126, \dodoi{10.1111/j.1365-2966.2005.09981.x}

\bibitem[{{Carnall} {et~al.}(2023){Carnall}, {McLure}, {Dunlop}, {McLeod}, {Wild}, {Cullen}, {Magee}, {Begley}, {Cimatti}, {Donnan}, {Hamadouche}, {Jewell}, \& {Walker}}]{Carnall2023_nature}
{Carnall}, A.~C., {McLure}, R.~J., {Dunlop}, J.~S., {et~al.} 2023, \nat, 619, 716, \dodoi{10.1038/s41586-023-06158-6}

\bibitem[{{Chabrier}(2003)}]{Chabrier2003}
{Chabrier}, G. 2003, \pasp, 115, 763, \dodoi{10.1086/376392}

\bibitem[{{Chen} {et~al.}(2023){Chen}, {Stark}, {Endsley}, {Topping}, {Whitler}, \& {Charlot}}]{Chen2023}
{Chen}, Z., {Stark}, D.~P., {Endsley}, R., {et~al.} 2023, \mnras, 518, 5607, \dodoi{10.1093/mnras/stac3476}

\bibitem[{{Conroy} {et~al.}(2009){Conroy}, {Gunn}, \& {White}}]{Conroy2009}
{Conroy}, C., {Gunn}, J.~E., \& {White}, M. 2009, \apj, 699, 486, \dodoi{10.1088/0004-637X/699/1/486}

\bibitem[{{Costantin} {et~al.}(2023){Costantin}, {P{\'e}rez-Gonz{\'a}lez}, {Vega-Ferrero}, {Huertas-Company}, {Bisigello}, {Buitrago}, {Bagley}, {Cleri}, {Cooper}, {Finkelstein}, {Holwerda}, {Kartaltepe}, {Koekemoer}, {Nelson}, {Papovich}, {Pillepich}, {Pirzkal}, {Tacchella}, \& {Yung}}]{Costantin2023}
{Costantin}, L., {P{\'e}rez-Gonz{\'a}lez}, P.~G., {Vega-Ferrero}, J., {et~al.} 2023, \apj, 946, 71, \dodoi{10.3847/1538-4357/acb926}

\bibitem[{{Davidson} \& {Kinman}(1985)}]{Davidson1985}
{Davidson}, K., \& {Kinman}, T.~D. 1985, \apjs, 58, 321, \dodoi{10.1086/191044}

\bibitem[{{de Graaff} {et~al.}(2024{\natexlab{a}}){de Graaff}, {Brammer}, {Weibel}, {Lewis}, {Maseda}, {Oesch}, {Bezanson}, {Boogaard}, {Cleri}, {Cooper}, {Gottumukkala}, {Greene}, {Hirschmann}, {Hviding}, {Katz}, {Labb{\'e}}, {Leja}, {Matthee}, {McConachie}, {Miller}, {Naidu}, {Price}, {Rix}, {Setton}, {Suess}, {Wang}, {Whitaker}, \& {Williams}}]{Graaff2024_rubies}
{de Graaff}, A., {Brammer}, G., {Weibel}, A., {et~al.} 2024{\natexlab{a}}, arXiv e-prints, arXiv:2409.05948, \dodoi{10.48550/arXiv.2409.05948}

\bibitem[{{de Graaff} {et~al.}(2024{\natexlab{b}}){de Graaff}, {Setton}, {Brammer}, {Cutler}, {Suess}, {Labbe}, {Leja}, {Weibel}, {Maseda}, {Whitaker}, {Bezanson}, {Boogaard}, {Cleri}, {De Lucia}, {Franx}, {Greene}, {Hirschmann}, {Matthee}, {McConachie}, {Naidu}, {Oesch}, {Price}, {Rix}, {Valentino}, {Wang}, \& {Williams}}]{Graaff2024}
{de Graaff}, A., {Setton}, D.~J., {Brammer}, G., {et~al.} 2024{\natexlab{b}}, arXiv e-prints, arXiv:2404.05683, \dodoi{10.48550/arXiv.2404.05683}

\bibitem[{{de Jong} \& {Lacey}(2000)}]{DeJong2000}
{de Jong}, R.~S., \& {Lacey}, C. 2000, \apj, 545, 781, \dodoi{10.1086/317840}

\bibitem[{{Dekel} {et~al.}(2023){Dekel}, {Sarkar}, {Birnboim}, {Mandelker}, \& {Li}}]{Dekel2023}
{Dekel}, A., {Sarkar}, K.~C., {Birnboim}, Y., {Mandelker}, N., \& {Li}, Z. 2023, \mnras, 523, 3201, \dodoi{10.1093/mnras/stad1557}

\bibitem[{{Dekel} {et~al.}(2024){Dekel}, {Stone}, {Dutta Chowdhury}, {Gilbaum}, {Li}, {Mandelker}, \& {van den Bosch}}]{Dekel2024}
{Dekel}, A., {Stone}, N.~C., {Dutta Chowdhury}, D., {et~al.} 2024, arXiv e-prints, arXiv:2409.18605, \dodoi{10.48550/arXiv.2409.18605}

\bibitem[{{Diamond-Stanic} {et~al.}(2021){Diamond-Stanic}, {Moustakas}, {Sell}, {Tremonti}, {Coil}, {Davis}, {Geach}, {Gottlieb}, {Hickox}, {Kepley}, {Lipscomb}, {Rines}, {Rudnick}, {Thompson}, {Valdez}, {Bradna}, {Camarillo}, {Cinquino}, {Ohene}, {Perrotta}, {Petter}, {Rupke}, {Umeh}, \& {Whalen}}]{Stanic2021}
{Diamond-Stanic}, A.~M., {Moustakas}, J., {Sell}, P.~H., {et~al.} 2021, \apj, 912, 11, \dodoi{10.3847/1538-4357/abe935}

\bibitem[{{Ding} {et~al.}(2022){Ding}, {Silverman}, \& {Onoue}}]{Ding2022}
{Ding}, X., {Silverman}, J.~D., \& {Onoue}, M. 2022, \apjl, 939, L28, \dodoi{10.3847/2041-8213/ac9c02}

\bibitem[{{Draine} \& {McKee}(1993)}]{Draine1993}
{Draine}, B.~T., \& {McKee}, C.~F. 1993, \araa, 31, 373, \dodoi{10.1146/annurev.aa.31.090193.002105}

\bibitem[{{Ebisuzaki} {et~al.}(1991){Ebisuzaki}, {Makino}, \& {Okumura}}]{Ebisuzaki1991}
{Ebisuzaki}, T., {Makino}, J., \& {Okumura}, S.~K. 1991, \nat, 354, 212, \dodoi{10.1038/354212a0}

\bibitem[{{Eisenstein} \& {Loeb}(1995)}]{EisensteinLoeb1995}
{Eisenstein}, D.~J., \& {Loeb}, A. 1995, \apj, 443, 11, \dodoi{10.1086/175498}

\bibitem[{{Fan} {et~al.}(2008){Fan}, {Lapi}, {De Zotti}, \& {Danese}}]{Fan2008}
{Fan}, L., {Lapi}, A., {De Zotti}, G., \& {Danese}, L. 2008, \apjl, 689, L101, \dodoi{10.1086/595784}

\bibitem[{{Finkelstein} {et~al.}(2023{\natexlab{a}}){Finkelstein}, {Bagley}, \& {Yang}}]{doiCEERSmosaic}
{Finkelstein}, S.~L., {Bagley}, M.~B., \& {Yang}, G. 2023{\natexlab{a}}, Data from The Cosmic Evolution Early Release Science Survey (CEERS),  STScI/MAST, \dodoi{10.17909/Z7P0-8481}

\bibitem[{{Finkelstein} {et~al.}(2022){Finkelstein}, {Bagley}, {Haro}, {Dickinson}, {Ferguson}, {Kartaltepe}, {Papovich}, {Burgarella}, {Kocevski}, {Huertas-Company}, {Iyer}, {Koekemoer}, {Larson}, {P{\'e}rez-Gonz{\'a}lez}, {Rose}, {Tacchella}, {Wilkins}, {Chworowsky}, {Medrano}, {Morales}, {Somerville}, {Yung}, {Fontana}, {Giavalisco}, {Grazian}, {Grogin}, {Kewley}, {Kirkpatrick}, {Kurczynski}, {Lotz}, {Pentericci}, {Pirzkal}, {Ravindranath}, {Ryan}, {Trump}, {Yang}, {Almaini}, {Amor{\'\i}n}, {Annunziatella}, {Backhaus}, {Barro}, {Behroozi}, {Bell}, {Bhatawdekar}, {Bisigello}, {Bromm}, {Buat}, {Buitrago}, {Calabr{\`o}}, {Casey}, {Castellano}, {Ch{\'a}vez Ortiz}, {Ciesla}, {Cleri}, {Cohen}, {Cole}, {Cooke}, {Cooper}, {Cooray}, {Costantin}, {Cox}, {Croton}, {Daddi}, {Dav{\'e}}, {de La Vega}, {Dekel}, {Elbaz}, {Estrada-Carpenter}, {Faber}, {Fern{\'a}ndez}, {Finkelstein}, {Freundlich}, {Fujimoto}, {Garc{\'\i}a-Argum{\'a}nez}, {Gardner}, {Gawiser}, {G{\'o}mez-Guijarro}, {Guo}, {Hamblin}, {Hamilton}, {Hathi},
  {Holwerda}, {Hirschmann}, {Hutchison}, {Jaskot}, {Jha}, {Jogee}, {Juneau}, {Jung}, {Kassin}, {Le Bail}, {Leung}, {Lucas}, {Magnelli}, {Mantha}, {Matharu}, {McGrath}, {McIntosh}, {Merlin}, {Mobasher}, {Newman}, {Nicholls}, {Pandya}, {Rafelski}, {Ronayne}, {Santini}, {Seill{\'e}}, {Shah}, {Shen}, {Simons}, {Snyder}, {Stanway}, {Straughn}, {Teplitz}, {Vanderhoof}, {Vega-Ferrero}, {Wang}, {Weiner}, {Willmer}, {Wuyts}, {Zavala}, \& {CEERS Team}}]{Finkelstein2022}
{Finkelstein}, S.~L., {Bagley}, M.~B., {Haro}, P.~A., {et~al.} 2022, \apjl, 940, L55, \dodoi{10.3847/2041-8213/ac966e}

\bibitem[{{Finkelstein} {et~al.}(2023{\natexlab{b}}){Finkelstein}, {Bagley}, {Ferguson}, {Wilkins}, {Kartaltepe}, {Papovich}, {Yung}, {Arrabal Haro}, {Behroozi}, {Dickinson}, {Kocevski}, {Koekemoer}, {Larson}, {Le Bail}, {Morales}, {P{\'e}rez-Gonz{\'a}lez}, {Burgarella}, {Dav{\'e}}, {Hirschmann}, {Somerville}, {Wuyts}, {Bromm}, {Casey}, {Fontana}, {Fujimoto}, {Gardner}, {Giavalisco}, {Grazian}, {Grogin}, {Hathi}, {Hutchison}, {Jha}, {Jogee}, {Kewley}, {Kirkpatrick}, {Long}, {Lotz}, {Pentericci}, {Pierel}, {Pirzkal}, {Ravindranath}, {Ryan}, {Trump}, {Yang}, {Bhatawdekar}, {Bisigello}, {Buat}, {Calabr{\`o}}, {Castellano}, {Cleri}, {Cooper}, {Croton}, {Daddi}, {Dekel}, {Elbaz}, {Franco}, {Gawiser}, {Holwerda}, {Huertas-Company}, {Jaskot}, {Leung}, {Lucas}, {Mobasher}, {Pandya}, {Tacchella}, {Weiner}, \& {Zavala}}]{Finkelstein2023}
{Finkelstein}, S.~L., {Bagley}, M.~B., {Ferguson}, H.~C., {et~al.} 2023{\natexlab{b}}, \apjl, 946, L13, \dodoi{10.3847/2041-8213/acade4}

\bibitem[{{Franx}(1993)}]{Franx1993}
{Franx}, M. 1993, in IAU Symposium, Vol. 153, Galactic Bulges, ed. H.~{Dejonghe} \& H.~J. {Habing}, 243

\bibitem[{{Franx} {et~al.}(2008){Franx}, {van Dokkum}, {F{\"o}rster Schreiber}, {Wuyts}, {Labb{\'e}}, \& {Toft}}]{Franx2008}
{Franx}, M., {van Dokkum}, P.~G., {F{\"o}rster Schreiber}, N.~M., {et~al.} 2008, \apj, 688, 770, \dodoi{10.1086/592431}

\bibitem[{{Furtak} {et~al.}(2023){Furtak}, {Zitrin}, {Plat}, {Fujimoto}, {Wang}, {Nelson}, {Labb{\'e}}, {Bezanson}, {Brammer}, {van Dokkum}, {Endsley}, {Glazebrook}, {Greene}, {Leja}, {Price}, {Smit}, {Stark}, {Weaver}, {Whitaker}, {Atek}, {Chevallard}, {Curtis-Lake}, {Dayal}, {Feltre}, {Franx}, {Fudamoto}, {Marchesini}, {Mowla}, {Pan}, {Suess}, {Vidal-Garc{\'\i}a}, \& {Williams}}]{Furtak2023b}
{Furtak}, L.~J., {Zitrin}, A., {Plat}, A., {et~al.} 2023, \apj, 952, 142, \dodoi{10.3847/1538-4357/acdc9d}

\bibitem[{{Furtak} {et~al.}(2024){Furtak}, {Labb{\'e}}, {Zitrin}, {Greene}, {Dayal}, {Chemerynska}, {Kokorev}, {Miller}, {Goulding}, {de Graaff}, {Bezanson}, {Brammer}, {Cutler}, {Leja}, {Pan}, {Price}, {Wang}, {Weaver}, {Whitaker}, {Atek}, {Bogd{\'a}n}, {Charlot}, {Curtis-Lake}, {van Dokkum}, {Endsley}, {Feldmann}, {Fudamoto}, {Fujimoto}, {Glazebrook}, {Juneau}, {Marchesini}, {Maseda}, {Nelson}, {Oesch}, {Plat}, {Setton}, {Stark}, \& {Williams}}]{Furtak2024}
{Furtak}, L.~J., {Labb{\'e}}, I., {Zitrin}, A., {et~al.} 2024, \nat, 628, 57, \dodoi{10.1038/s41586-024-07184-8}

\bibitem[{{Grazian} {et~al.}(2012){Grazian}, {Castellano}, {Fontana}, {Pentericci}, {Dunlop}, {McLure}, {Koekemoer}, {Dickinson}, {Faber}, {Ferguson}, {Galametz}, {Giavalisco}, {Grogin}, {Hathi}, {Kocevski}, {Lai}, {Newman}, \& {Vanzella}}]{Grazian2012}
{Grazian}, A., {Castellano}, M., {Fontana}, A., {et~al.} 2012, \aap, 547, A51, \dodoi{10.1051/0004-6361/201219669}

\bibitem[{{Greene} {et~al.}(2024){Greene}, {Labbe}, {Goulding}, {Furtak}, {Chemerynska}, {Kokorev}, {Dayal}, {Volonteri}, {Williams}, {Wang}, {Setton}, {Burgasser}, {Bezanson}, {Atek}, {Brammer}, {Cutler}, {Feldmann}, {Fujimoto}, {Glazebrook}, {de Graaff}, {Khullar}, {Leja}, {Marchesini}, {Maseda}, {Matthee}, {Miller}, {Naidu}, {Nanayakkara}, {Oesch}, {Pan}, {Papovich}, {Price}, {van Dokkum}, {Weaver}, {Whitaker}, \& {Zitrin}}]{Greene2024}
{Greene}, J.~E., {Labbe}, I., {Goulding}, A.~D., {et~al.} 2024, \apj, 964, 39, \dodoi{10.3847/1538-4357/ad1e5f}

\bibitem[{{Grudi{\'c}} {et~al.}(2019){Grudi{\'c}}, {Hopkins}, {Quataert}, \& {Murray}}]{Grudic2019}
{Grudi{\'c}}, M.~Y., {Hopkins}, P.~F., {Quataert}, E., \& {Murray}, N. 2019, \mnras, 483, 5548, \dodoi{10.1093/mnras/sty3386}

\bibitem[{{Hamilton}(1985)}]{Hamilton1985}
{Hamilton}, D. 1985, \apj, 297, 371, \dodoi{10.1086/163537}

\bibitem[{{Harikane} {et~al.}(2023{\natexlab{a}}){Harikane}, {Zhang}, {Nakajima}, {Ouchi}, {Isobe}, {Ono}, {Hatano}, {Xu}, \& {Umeda}}]{Harikane2023}
{Harikane}, Y., {Zhang}, Y., {Nakajima}, K., {et~al.} 2023{\natexlab{a}}, \apj, 959, 39, \dodoi{10.3847/1538-4357/ad029e}

\bibitem[{{Harikane} {et~al.}(2023{\natexlab{b}}){Harikane}, {Ouchi}, {Oguri}, {Ono}, {Nakajima}, {Isobe}, {Umeda}, {Mawatari}, \& {Zhang}}]{Harikane2023_imf}
{Harikane}, Y., {Ouchi}, M., {Oguri}, M., {et~al.} 2023{\natexlab{b}}, \apjs, 265, 5, \dodoi{10.3847/1538-4365/acaaa9}

\bibitem[{{Heintz} {et~al.}(2024){Heintz}, {Brammer}, {Watson}, {Oesch}, {Keating}, {Hayes}, {Abdurro'uf}, {Arellano-C{\'o}rdova}, {Carnall}, {Christiansen}, {Cullen}, {Dav{\'e}}, {Dayal}, {Ferrara}, {Finlator}, {Fynbo}, {Flury}, {Gelli}, {Gillman}, {Gottumukkala}, {Gould}, {Greve}, {Hardin}, {Y. -Y Hsiao}, {Hutter}, {Jakobsson}, {Killi}, {Khosravaninezhad}, {Laursen}, {Lee}, {Magdis}, {Matthee}, {Naidu}, {Narayanan}, {Pollock}, {Prescott}, {Rusakov}, {Shuntov}, {Sneppen}, {Smit}, {Tanvir}, {Terp}, {Toft}, {Valentino}, {Vijayan}, {Weaver}, {Wise}, \& {Witstok}}]{Heintz2024}
{Heintz}, K.~E., {Brammer}, G.~B., {Watson}, D., {et~al.} 2024, arXiv e-prints, arXiv:2404.02211, \dodoi{10.48550/arXiv.2404.02211}

\bibitem[{{Hills}(1983)}]{Hills1983}
{Hills}, J.~G. 1983, \aj, 88, 1269, \dodoi{10.1086/113418}

\bibitem[{{Hopkins} {et~al.}(2010){Hopkins}, {Murray}, {Quataert}, \& {Thompson}}]{Hopkins2010_density}
{Hopkins}, P.~F., {Murray}, N., {Quataert}, E., \& {Thompson}, T.~A. 2010, \mnras, 401, L19, \dodoi{10.1111/j.1745-3933.2009.00777.x}

\bibitem[{{Iani} {et~al.}(2024){Iani}, {Rinaldi}, {Caputi}, {Annunziatella}, {Langeroodi}, {Melinder}, {P{\'e}rez-Gonz{\'a}lez}, {{\'A}lvarez-M{\'a}rquez}, {Boogaard}, {Bosman}, {Costantin}, {Moutard}, {Colina}, {{\"O}stlin}, {Greve}, {Wright}, {Alonso-Herrero}, {Bik}, {Gillman}, {Crespo G{\'o}mez}, {Hjorth}, {Labiano}, {Pye}, {Tikkanen}, \& {van der Werf}}]{Iani2024}
{Iani}, E., {Rinaldi}, P., {Caputi}, K.~I., {et~al.} 2024, arXiv e-prints, arXiv:2406.18207, \dodoi{10.48550/arXiv.2406.18207}

\bibitem[{{Inayoshi} {et~al.}(2022){Inayoshi}, {Harikane}, {Inoue}, {Li}, \& {Ho}}]{inayoshi2022}
{Inayoshi}, K., {Harikane}, Y., {Inoue}, A.~K., {Li}, W., \& {Ho}, L.~C. 2022, \apjl, 938, L10, \dodoi{10.3847/2041-8213/ac9310}

\bibitem[{{Inayoshi} \& {Ichikawa}(2024)}]{Inayoshi2024}
{Inayoshi}, K., \& {Ichikawa}, K. 2024, \apjl, 973, L49, \dodoi{10.3847/2041-8213/ad74e2}

\bibitem[{{Ji} {et~al.}(2024){Ji}, {Williams}, {Suess}, {Tacchella}, {Johnson}, {Robertson}, {Alberts}, {Baker}, {Baum}, {Bhatawdekar}, {Bonaventura}, {Boyett}, {Bunker}, {Carniani}, {Charlot}, {Chen}, {Chevallard}, {Curtis-Lake}, {D'Eugenio}, {de Graaff}, {DeCoursey}, {Egami}, {Eisenstein}, {Hainline}, {Hausen}, {Helton}, {Looser}, {Lyu}, {Maiolino}, {Maseda}, {Nelson}, {Rieke}, {Rieke}, {Rix}, {Sandles}, {Sun}, {{\"U}bler}, {Willmer}, {Willott}, \& {Witstok}}]{Ji2024}
{Ji}, Z., {Williams}, C.~C., {Suess}, K.~A., {et~al.} 2024, arXiv e-prints, arXiv:2401.00934, \dodoi{10.48550/arXiv.2401.00934}

\bibitem[{{Juod{\v{z}}balis} {et~al.}(2024{\natexlab{a}}){Juod{\v{z}}balis}, {Maiolino}, {Baker}, {Tacchella}, {Scholtz}, {D'Eugenio}, {Schneider}, {Trinca}, {Valiante}, {DeCoursey}, {Curti}, {Carniani}, {Chevallard}, {de Graaff}, {Arribas}, {Bennett}, {Bourne}, {Bunker}, {Charlot}, {Jiang}, {Koudmani}, {Perna}, {Robertson}, {Sijacki}, {{\"U}bler}, {Williams}, {Willott}, \& {Witstok}}]{Juodzbalis2024}
{Juod{\v{z}}balis}, I., {Maiolino}, R., {Baker}, W.~M., {et~al.} 2024{\natexlab{a}}, arXiv e-prints, arXiv:2403.03872, \dodoi{10.48550/arXiv.2403.03872}

\bibitem[{{Juod{\v{z}}balis} {et~al.}(2024{\natexlab{b}}){Juod{\v{z}}balis}, {Ji}, {Maiolino}, {D'Eugenio}, {Scholtz}, {Risaliti}, {Fabian}, {Mazzolari}, {Gilli}, {Prandoni}, {Arribas}, {Bunker}, {Carniani}, {Charlot}, {Curtis-Lake}, {de Graaff}, {Hainline}, {Parlanti}, {Perna}, {P{\'e}rez-Gonz{\'a}lez}, {Robertson}, {Tacchella}, {{\"U}bler}, {Williams}, {Willott}, \& {Witstok}}]{Juodzbalis2024_rosetta}
{Juod{\v{z}}balis}, I., {Ji}, X., {Maiolino}, R., {et~al.} 2024{\natexlab{b}}, \mnras, 535, 853, \dodoi{10.1093/mnras/stae2367}

\bibitem[{{Kawamata} {et~al.}(2018){Kawamata}, {Ishigaki}, {Shimasaku}, {Oguri}, {Ouchi}, \& {Tanigawa}}]{Kawamata2018}
{Kawamata}, R., {Ishigaki}, M., {Shimasaku}, K., {et~al.} 2018, \apj, 855, 4, \dodoi{10.3847/1538-4357/aaa6cf}

\bibitem[{{Killi} {et~al.}(2024){Killi}, {Watson}, {Brammer}, {McPartland}, {Antwi-Danso}, {Newshore}, {Coe}, {Allen}, {Fynbo}, {Gould}, {Heintz}, {Rusakov}, \& {Vejlgaard}}]{Killi2024}
{Killi}, M., {Watson}, D., {Brammer}, G., {et~al.} 2024, \aap, 691, A52, \dodoi{10.1051/0004-6361/202348857}

\bibitem[{{Kocevski} {et~al.}(2023){Kocevski}, {Onoue}, {Inayoshi}, {Trump}, {Arrabal Haro}, {Grazian}, {Dickinson}, {Finkelstein}, {Kartaltepe}, {Hirschmann}, {Aird}, {Holwerda}, {Fujimoto}, {Juneau}, {Amor{\'\i}n}, {Backhaus}, {Bagley}, {Barro}, {Bell}, {Bisigello}, {Calabr{\`o}}, {Cleri}, {Cooper}, {Ding}, {Grogin}, {Ho}, {Hutchison}, {Inoue}, {Jiang}, {Jones}, {Koekemoer}, {Li}, {Li}, {McGrath}, {Molina}, {Papovich}, {P{\'e}rez-Gonz{\'a}lez}, {Pirzkal}, {Wilkins}, {Yang}, \& {Yung}}]{Kocevski2023}
{Kocevski}, D.~D., {Onoue}, M., {Inayoshi}, K., {et~al.} 2023, \apjl, 954, L4, \dodoi{10.3847/2041-8213/ace5a0}

\bibitem[{{Kocevski} {et~al.}(2024){Kocevski}, {Finkelstein}, {Barro}, {Taylor}, {Calabr{\`o}}, {Laloux}, {Buchner}, {Trump}, {Leung}, {Yang}, {Dickinson}, {P{\'e}rez-Gonz{\'a}lez}, {Pacucci}, {Inayoshi}, {Somerville}, {McGrath}, {Akins}, {Bagley}, {Bisigello}, {Bowler}, {Carnall}, {Casey}, {Cheng}, {Cleri}, {Costantin}, {Cullen}, {Davis}, {Donnan}, {Dunlop}, {Ellis}, {Ferguson}, {Fujimoto}, {Fontana}, {Giavalisco}, {Grazian}, {Grogin}, {Hathi}, {Hirschmann}, {Huertas-Company}, {Holwerda}, {Illingworth}, {Juneau}, {Kartaltepe}, {Koekemoer}, {Li}, {Lucas}, {Magee}, {Mason}, {McLeod}, {McLure}, {Napolitano}, {Papovich}, {Pirzkal}, {Rodighiero}, {Santini}, {Wilkins}, \& {Yung}}]{Kocevski2024}
{Kocevski}, D.~D., {Finkelstein}, S.~L., {Barro}, G., {et~al.} 2024, arXiv e-prints, arXiv:2404.03576, \dodoi{10.48550/arXiv.2404.03576}

\bibitem[{{Kokorev} {et~al.}(2023){Kokorev}, {Fujimoto}, {Labbe}, {Greene}, {Bezanson}, {Dayal}, {Nelson}, {Atek}, {Brammer}, {Caputi}, {Chemerynska}, {Cutler}, {Feldmann}, {Fudamoto}, {Furtak}, {Goulding}, {de Graaff}, {Leja}, {Marchesini}, {Miller}, {Nanayakkara}, {Oesch}, {Pan}, {Price}, {Setton}, {Smit}, {Stefanon}, {Wang}, {Weaver}, {Whitaker}, {Williams}, \& {Zitrin}}]{Kokorev2023}
{Kokorev}, V., {Fujimoto}, S., {Labbe}, I., {et~al.} 2023, \apjl, 957, L7, \dodoi{10.3847/2041-8213/ad037a}

\bibitem[{{Kokorev} {et~al.}(2024){Kokorev}, {Chisholm}, {Endsley}, {Finkelstein}, {Greene}, {Akins}, {Bromm}, {Casey}, {Fujimoto}, {Labb{\'e}}, \& {Larson}}]{Kokorev2024}
{Kokorev}, V., {Chisholm}, J., {Endsley}, R., {et~al.} 2024, \apj, 975, 178, \dodoi{10.3847/1538-4357/ad7d03}

\bibitem[{{Kokubo} \& {Harikane}(2024)}]{Kokubo2024}
{Kokubo}, M., \& {Harikane}, Y. 2024, arXiv e-prints, arXiv:2407.04777.
\newblock \doarXiv{2407.04777}

\bibitem[{{Kov{\'a}cs} {et~al.}(2024){Kov{\'a}cs}, {Bogd{\'a}n}, {Natarajan}, {Werner}, {Azadi}, {Volonteri}, {Tremblay}, {Chadayammuri}, {Forman}, {Jones}, \& {Kraft}}]{Kovacs2024}
{Kov{\'a}cs}, O.~E., {Bogd{\'a}n}, {\'A}., {Natarajan}, P., {et~al.} 2024, \apjl, 965, L21, \dodoi{10.3847/2041-8213/ad391f}

\bibitem[{{Kroupa} {et~al.}(2020){Kroupa}, {Subr}, {Jerabkova}, \& {Wang}}]{Kroupa2020}
{Kroupa}, P., {Subr}, L., {Jerabkova}, T., \& {Wang}, L. 2020, \mnras, 498, 5652, \dodoi{10.1093/mnras/staa2276}

\bibitem[{{Labb{\'e}} {et~al.}(2023{\natexlab{a}}){Labb{\'e}}, {van Dokkum}, {Nelson}, {Bezanson}, {Suess}, {Leja}, {Brammer}, {Whitaker}, {Mathews}, {Stefanon}, \& {Wang}}]{Labbe2023}
{Labb{\'e}}, I., {van Dokkum}, P., {Nelson}, E., {et~al.} 2023{\natexlab{a}}, \nat, 616, 266, \dodoi{10.1038/s41586-023-05786-2}

\bibitem[{{Labb{\'e}} {et~al.}(2023{\natexlab{b}}){Labb{\'e}}, {Greene}, {Bezanson}, {Fujimoto}, {Furtak}, {Goulding}, {Matthee}, {Naidu}, {Oesch}, {Atek}, {Brammer}, {Chemerynska}, {Coe}, {Cutler}, {Dayal}, {Feldmann}, {Franx}, {Glazebrook}, {Leja}, {Marchesini}, {Maseda}, {Nanayakkara}, {Nelson}, {Pan}, {Papovich}, {Price}, {Suess}, {Wang}, {Whitaker}, {Williams}, \& {Zitrin}}]{Labbe2023b}
{Labb{\'e}}, I., {Greene}, J.~E., {Bezanson}, R., {et~al.} 2023{\natexlab{b}}, arXiv e-prints, arXiv:2306.07320, \dodoi{10.48550/arXiv.2306.07320}

\bibitem[{{Langeroodi} \& {Hjorth}(2023)}]{Langeroodi2023}
{Langeroodi}, D., \& {Hjorth}, J. 2023, arXiv e-prints, arXiv:2307.06336, \dodoi{10.48550/arXiv.2307.06336}

\bibitem[{{Larson} {et~al.}(2023){Larson}, {Finkelstein}, {Kocevski}, {Hutchison}, {Trump}, {Arrabal Haro}, {Bromm}, {Cleri}, {Dickinson}, {Fujimoto}, {Kartaltepe}, {Koekemoer}, {Papovich}, {Pirzkal}, {Tacchella}, {Zavala}, {Bagley}, {Behroozi}, {Champagne}, {Cole}, {Jung}, {Morales}, {Yang}, {Zhang}, {Zitrin}, {Amor{\'\i}n}, {Burgarella}, {Casey}, {Ch{\'a}vez Ortiz}, {Cox}, {Chworowsky}, {Fontana}, {Gawiser}, {Grazian}, {Grogin}, {Harish}, {Hathi}, {Hirschmann}, {Holwerda}, {Juneau}, {Leung}, {Lucas}, {McGrath}, {P{\'e}rez-Gonz{\'a}lez}, {Rigby}, {Seill{\'e}}, {Simons}, {de La Vega}, {Weiner}, {Wilkins}, {Yung}, \& {Ceers Team}}]{Larson2023}
{Larson}, R.~L., {Finkelstein}, S.~L., {Kocevski}, D.~D., {et~al.} 2023, \apjl, 953, L29, \dodoi{10.3847/2041-8213/ace619}

\bibitem[{{Li} {et~al.}(2024){Li}, {Dekel}, {Sarkar}, {Aung}, {Giavalisco}, {Mandelker}, \& {Tacchella}}]{Li2024}
{Li}, Z., {Dekel}, A., {Sarkar}, K.~C., {et~al.} 2024, \aap, 690, A108, \dodoi{10.1051/0004-6361/202348727}

\bibitem[{{Loeb}(2024)}]{Loeb2024}
{Loeb}, A. 2024, Research Notes of the American Astronomical Society, 8, 182, \dodoi{10.3847/2515-5172/ad614c}

\bibitem[{{Lynden-Bell} \& {Wood}(1968)}]{LyndenbellWood1968}
{Lynden-Bell}, D., \& {Wood}, R. 1968, \mnras, 138, 495, \dodoi{10.1093/mnras/138.4.495}

\bibitem[{{Ma} {et~al.}(2024){Ma}, {Greene}, {Setton}, {Volonteri}, {Leja}, {Wang}, {Bezanson}, {Brammer}, {Cutler}, {Dayal}, {van Dokkum}, {Furtak}, {Glazebrook}, {Goulding}, {de Graaff}, {Kokorev}, {Labbe}, {Pan}, {Price}, {Weaver}, {Williams}, {Whitaker}, \& {Zitrin}}]{Ma2024}
{Ma}, Y., {Greene}, J.~E., {Setton}, D.~J., {et~al.} 2024, arXiv e-prints, arXiv:2410.06257, \dodoi{10.48550/arXiv.2410.06257}

\bibitem[{{Maiolino} {et~al.}(2024{\natexlab{a}}){Maiolino}, {Scholtz}, {Curtis-Lake}, {Carniani}, {Baker}, {de Graaff}, {Tacchella}, {{\"U}bler}, {D'Eugenio}, {Witstok}, {Curti}, {Arribas}, {Bunker}, {Charlot}, {Chevallard}, {Eisenstein}, {Egami}, {Ji}, {Jones}, {Lyu}, {Rawle}, {Robertson}, {Rujopakarn}, {Perna}, {Sun}, {Venturi}, {Williams}, \& {Willott}}]{Maiolino2024_jades}
{Maiolino}, R., {Scholtz}, J., {Curtis-Lake}, E., {et~al.} 2024{\natexlab{a}}, \aap, 691, A145, \dodoi{10.1051/0004-6361/202347640}

\bibitem[{{Maiolino} {et~al.}(2024{\natexlab{b}}){Maiolino}, {Risaliti}, {Signorini}, {Trefoloni}, {Juodzbalis}, {Scholtz}, {Uebler}, {D'Eugenio}, {Carniani}, {Fabian}, {Ji}, {Mazzolari}, {Bertola}, {Brusa}, {Bunker}, {Charlot}, {Comastri}, {Cresci}, {DeCoursey}, {Egami}, {Fiore}, {Gilli}, {Perna}, {Tacchella}, \& {Venturi}}]{Maiolino2024}
{Maiolino}, R., {Risaliti}, G., {Signorini}, M., {et~al.} 2024{\natexlab{b}}, arXiv e-prints, arXiv:2405.00504, \dodoi{10.48550/arXiv.2405.00504}

\bibitem[{{Marshall} {et~al.}(2022){Marshall}, {Wilkins}, {Di Matteo}, {Roper}, {Vijayan}, {Ni}, {Feng}, \& {Croft}}]{Marshall2022}
{Marshall}, M.~A., {Wilkins}, S., {Di Matteo}, T., {et~al.} 2022, \mnras, 511, 5475, \dodoi{10.1093/mnras/stac380}

\bibitem[{{Matthee} {et~al.}(2024){Matthee}, {Naidu}, {Brammer}, {Chisholm}, {Eilers}, {Goulding}, {Greene}, {Kashino}, {Labbe}, {Lilly}, {Mackenzie}, {Oesch}, {Weibel}, {Wuyts}, {Xiao}, {Bordoloi}, {Bouwens}, {van Dokkum}, {Illingworth}, {Kramarenko}, {Maseda}, {Mason}, {Meyer}, {Nelson}, {Reddy}, {Shivaei}, {Simcoe}, \& {Yue}}]{Matthee2024}
{Matthee}, J., {Naidu}, R.~P., {Brammer}, G., {et~al.} 2024, \apj, 963, 129, \dodoi{10.3847/1538-4357/ad2345}

\bibitem[{{Menon} {et~al.}(2024){Menon}, {Lancaster}, {Burkhart}, {Somerville}, {Dekel}, \& {Krumholz}}]{Menon2024}
{Menon}, S.~H., {Lancaster}, L., {Burkhart}, B., {et~al.} 2024, \apjl, 967, L28, \dodoi{10.3847/2041-8213/ad462d}

\bibitem[{{Naab} {et~al.}(2009){Naab}, {Johansson}, \& {Ostriker}}]{Naab2009}
{Naab}, T., {Johansson}, P.~H., \& {Ostriker}, J.~P. 2009, \apjl, 699, L178, \dodoi{10.1088/0004-637X/699/2/L178}

\bibitem[{Neumayer(2017)}]{Neumayer2017}
Neumayer, N. 2017, Proceedings of the International Astronomical Union, 12, 84–90, \dodoi{10.1017/S1743921316007018}

\bibitem[{{Ono} {et~al.}(2023){Ono}, {Harikane}, {Ouchi}, {Yajima}, {Abe}, {Isobe}, {Shibuya}, {Wise}, {Zhang}, {Nakajima}, \& {Umeda}}]{Ono2022}
{Ono}, Y., {Harikane}, Y., {Ouchi}, M., {et~al.} 2023, \apj, 951, 72, \dodoi{10.3847/1538-4357/acd44a}

\bibitem[{{Onoue} {et~al.}(2023){Onoue}, {Inayoshi}, {Ding}, {Li}, {Li}, {Molina}, {Inoue}, {Jiang}, \& {Ho}}]{Onoue2023}
{Onoue}, M., {Inayoshi}, K., {Ding}, X., {et~al.} 2023, \apjl, 942, L17, \dodoi{10.3847/2041-8213/aca9d3}

\bibitem[{{Pechetti} {et~al.}(2020){Pechetti}, {Seth}, {Neumayer}, {Georgiev}, {Kacharov}, \& {den Brok}}]{Pechetti2020}
{Pechetti}, R., {Seth}, A., {Neumayer}, N., {et~al.} 2020, \apj, 900, 32, \dodoi{10.3847/1538-4357/abaaa7}

\bibitem[{{Peng} {et~al.}(2002){Peng}, {Ho}, {Impey}, \& {Rix}}]{Peng2002}
{Peng}, C.~Y., {Ho}, L.~C., {Impey}, C.~D., \& {Rix}, H.-W. 2002, \aj, 124, 266, \dodoi{10.1086/340952}

\bibitem[{{Peng} {et~al.}(2010){Peng}, {Ho}, {Impey}, \& {Rix}}]{Peng2010x}
---. 2010, \aj, 139, 2097, \dodoi{10.1088/0004-6256/139/6/2097}

\bibitem[{{P{\'e}rez-Gonz{\'a}lez} {et~al.}(2024){P{\'e}rez-Gonz{\'a}lez}, {Barro}, {Rieke}, {Lyu}, {Rieke}, {Alberts}, {Williams}, {Hainline}, {Sun}, {Pusk{\'a}s}, {Annunziatella}, {Baker}, {Bunker}, {Egami}, {Ji}, {Johnson}, {Robertson}, {Rodr{\'\i}guez Del Pino}, {Rujopakarn}, {Shivaei}, {Tacchella}, {Willmer}, \& {Willott}}]{PerezGonzalez2024A}
{P{\'e}rez-Gonz{\'a}lez}, P.~G., {Barro}, G., {Rieke}, G.~H., {et~al.} 2024, \apj, 968, 4, \dodoi{10.3847/1538-4357/ad38bb}

\bibitem[{{Perrin} {et~al.}(2014){Perrin}, {Sivaramakrishnan}, {Lajoie}, {Elliott}, {Pueyo}, {Ravindranath}, \& {Albert}}]{Perrin2014}
{Perrin}, M.~D., {Sivaramakrishnan}, A., {Lajoie}, C.-P., {et~al.} 2014, {Proc SPIE}, 9143, 91433X, \dodoi{10.1117/12.2056689}

\bibitem[{{Poggianti} {et~al.}(1999){Poggianti}, {Smail}, {Dressler}, {Couch}, {Barger}, {Butcher}, {Ellis}, \& {Oemler}}]{Poggianti1999}
{Poggianti}, B.~M., {Smail}, I., {Dressler}, A., {et~al.} 1999, \apj, 518, 576, \dodoi{10.1086/307322}

\bibitem[{{Quinlan}(1996)}]{Quinlan1996}
{Quinlan}, G.~D. 1996, \na, 1, 35, \dodoi{10.1016/S1384-1076(96)00003-6}

\bibitem[{{Raga} {et~al.}(2015){Raga}, {Castellanos-Ram{\'\i}rez}, {Esquivel}, {Rodr{\'\i}guez-Gonz{\'a}lez}, \& {Vel{\'a}zquez}}]{Raga2015}
{Raga}, A.~C., {Castellanos-Ram{\'\i}rez}, A., {Esquivel}, A., {Rodr{\'\i}guez-Gonz{\'a}lez}, A., \& {Vel{\'a}zquez}, P.~F. 2015, \rmxaa, 51, 231

\bibitem[{{Rantala} {et~al.}(2024){Rantala}, {Rawlings}, {Naab}, {Thomas}, \& {Johansson}}]{Rantala2024}
{Rantala}, A., {Rawlings}, A., {Naab}, T., {Thomas}, J., \& {Johansson}, P.~H. 2024, \mnras, 535, 1202, \dodoi{10.1093/mnras/stae2424}

\bibitem[{Rix {et~al.}(1997)Rix, Guhathakurta, Colless, \& Ing}]{Rix1996}
Rix, H.~W., Guhathakurta, P., Colless, M., \& Ing, K. 1997, Mon. Not. Roy. Astron. Soc., 285, 779, \dodoi{10.1093/mnras/285.4.779}

\bibitem[{{Roper} {et~al.}(2023){Roper}, {Lovell}, {Vijayan}, {Irodotou}, {Kuusisto}, {Matharu}, {Seeyave}, {Thomas}, \& {Wilkins}}]{Roper2023}
{Roper}, W.~J., {Lovell}, C.~C., {Vijayan}, A.~P., {et~al.} 2023, \mnras, 526, 6128, \dodoi{10.1093/mnras/stad2746}

\bibitem[{{Rowland} {et~al.}(2024){Rowland}, {Hodge}, {Bouwens}, {Pi{\~n}a}, {Hygate}, {Algera}, {Aravena}, {Bowler}, {da Cunha}, {Dayal}, {Ferrara}, {Herard-Demanche}, {Inami}, {van Leeuwen}, {de Looze}, {Oesch}, {Pallottini}, {Phillips}, {Rybak}, {Schouws}, {Smit}, {Sommovigo}, {Stefanon}, \& {van der Werf}}]{Rowland2024}
{Rowland}, L.~E., {Hodge}, J., {Bouwens}, R., {et~al.} 2024, \mnras, \dodoi{10.1093/mnras/stae2217}

\bibitem[{{Schaerer} {et~al.}(2024{\natexlab{a}}){Schaerer}, {Guibert}, {Marques-Chaves}, \& {Martins}}]{Schaerer2024}
{Schaerer}, D., {Guibert}, J., {Marques-Chaves}, R., \& {Martins}, F. 2024{\natexlab{a}}, arXiv e-prints, arXiv:2407.12122, \dodoi{10.48550/arXiv.2407.12122}

\bibitem[{{Schaerer} {et~al.}(2024{\natexlab{b}}){Schaerer}, {Marques-Chaves}, {Xiao}, \& {Korber}}]{Schaerer2024_compact}
{Schaerer}, D., {Marques-Chaves}, R., {Xiao}, M., \& {Korber}, D. 2024{\natexlab{b}}, \aap, 687, L11, \dodoi{10.1051/0004-6361/202450721}

\bibitem[{{Sch{\"o}del} {et~al.}(2009){Sch{\"o}del}, {Merritt}, \& {Eckart}}]{Schodel2009}
{Sch{\"o}del}, R., {Merritt}, D., \& {Eckart}, A. 2009, \aap, 502, 91, \dodoi{10.1051/0004-6361/200810922}

\bibitem[{{Sersic}(1968)}]{Sersic1968}
{Sersic}, J.~L. 1968, {Atlas de Galaxias Australes}

\bibitem[{{Setton} {et~al.}(2024){Setton}, {Khullar}, {Miller}, {Bezanson}, {Greene}, {Suess}, {Whitaker}, {Antwi-Danso}, {Atek}, {Brammer}, {Cutler}, {Dayal}, {Feldmann}, {Fujimoto}, {Furtak}, {Glazebrook}, {Goulding}, {Kokorev}, {Labbe}, {Leja}, {Ma}, {Marchesini}, {Nanayakkara}, {Pan}, {Price}, {Siegel}, {Shipley}, {Weaver}, {van Dokkum}, {Wang}, \& {Williams}}]{Setton2024}
{Setton}, D.~J., {Khullar}, G., {Miller}, T.~B., {et~al.} 2024, \apj, 974, 145, \dodoi{10.3847/1538-4357/ad6a18}

\bibitem[{{Shen} {et~al.}(2024){Shen}, {Vogelsberger}, {Borrow}, {Hu}, {Erickson}, {Kannan}, {Smith}, {Garaldi}, {Hernquist}, {Morishita}, {Tacchella}, {Zier}, {Sun}, {Eilers}, \& {Wang}}]{Shen2024}
{Shen}, X., {Vogelsberger}, M., {Borrow}, J., {et~al.} 2024, \mnras, 534, 1433, \dodoi{10.1093/mnras/stae2156}

\bibitem[{{Shibuya} {et~al.}(2015){Shibuya}, {Ouchi}, \& {Harikane}}]{Shibuya2015}
{Shibuya}, T., {Ouchi}, M., \& {Harikane}, Y. 2015, \apjs, 219, 15, \dodoi{10.1088/0067-0049/219/2/15}

\bibitem[{{Silk} {et~al.}(2024){Silk}, {Begelman}, {Norman}, {Nusser}, \& {Wyse}}]{Silk2024}
{Silk}, J., {Begelman}, M.~C., {Norman}, C., {Nusser}, A., \& {Wyse}, R. F.~G. 2024, \apjl, 961, L39, \dodoi{10.3847/2041-8213/ad1bf0}

\bibitem[{{Stasi{\'n}ska} \& {Izotov}(2001)}]{Stasinska2001}
{Stasi{\'n}ska}, G., \& {Izotov}, Y. 2001, \aap, 378, 817, \dodoi{10.1051/0004-6361:20011303}

\bibitem[{{Stasi{\'n}ska} \& {Schaerer}(1999)}]{StasisnkaSchaerer1999}
{Stasi{\'n}ska}, G., \& {Schaerer}, D. 1999, \aap, 351, 72, \dodoi{10.48550/arXiv.astro-ph/9909203}

\bibitem[{{Tal} {et~al.}(2009){Tal}, {van Dokkum}, {Nelan}, \& {Bezanson}}]{Tal2009}
{Tal}, T., {van Dokkum}, P.~G., {Nelan}, J., \& {Bezanson}, R. 2009, \aj, 138, 1417, \dodoi{10.1088/0004-6256/138/5/1417}

\bibitem[{{Taylor} {et~al.}(2010){Taylor}, {Franx}, {Brinchmann}, {van der Wel}, \& {van Dokkum}}]{Taylor2010}
{Taylor}, E.~N., {Franx}, M., {Brinchmann}, J., {van der Wel}, A., \& {van Dokkum}, P.~G. 2010, \apj, 722, 1, \dodoi{10.1088/0004-637X/722/1/1}

\bibitem[{{Topping} {et~al.}(2024){Topping}, {Stark}, {Senchyna}, {Plat}, {Zitrin}, {Endsley}, {Charlot}, {Furtak}, {Maseda}, {Smit}, {Mainali}, {Chevallard}, {Molyneux}, \& {Rigby}}]{Topping2024}
{Topping}, M.~W., {Stark}, D.~P., {Senchyna}, P., {et~al.} 2024, \mnras, 529, 3301, \dodoi{10.1093/mnras/stae682}

\bibitem[{{Trinca} {et~al.}(2024){Trinca}, {Schneider}, {Valiante}, {Graziani}, {Ferrotti}, {Omukai}, \& {Chon}}]{Trinca2024}
{Trinca}, A., {Schneider}, R., {Valiante}, R., {et~al.} 2024, \mnras, 529, 3563, \dodoi{10.1093/mnras/stae651}

\bibitem[{{{\"U}bler} {et~al.}(2023){{\"U}bler}, {Maiolino}, {Curtis-Lake}, {P{\'e}rez-Gonz{\'a}lez}, {Curti}, {Perna}, {Arribas}, {Charlot}, {Marshall}, {D'Eugenio}, {Scholtz}, {Bunker}, {Carniani}, {Ferruit}, {Jakobsen}, {Rix}, {Rodr{\'\i}guez Del Pino}, {Willott}, {Boeker}, {Cresci}, {Jones}, {Kumari}, \& {Rawle}}]{Ubler2023}
{{\"U}bler}, H., {Maiolino}, R., {Curtis-Lake}, E., {et~al.} 2023, \aap, 677, A145, \dodoi{10.1051/0004-6361/202346137}

\bibitem[{{Upadhyaya} {et~al.}(2024){Upadhyaya}, {Marques-Chaves}, {Schaerer}, {Martins}, {P{\'e}rez-Fournon}, {Palacios}, \& {Stanway}}]{Upadhaya2024}
{Upadhyaya}, A., {Marques-Chaves}, R., {Schaerer}, D., {et~al.} 2024, \aap, 686, A185, \dodoi{10.1051/0004-6361/202449184}

\bibitem[{{van der Wel} {et~al.}(2006){van der Wel}, {Franx}, {Wuyts}, {van Dokkum}, {Huang}, {Rix}, \& {Illingworth}}]{vanderWel2006}
{van der Wel}, A., {Franx}, M., {Wuyts}, S., {et~al.} 2006, \apj, 652, 97, \dodoi{10.1086/508128}

\bibitem[{{van Dokkum} \& {Conroy}(2024)}]{Dokkum2024}
{van Dokkum}, P., \& {Conroy}, C. 2024, \apjl, 973, L32, \dodoi{10.3847/2041-8213/ad77b8}

\bibitem[{{van Dokkum} {et~al.}(2010){van Dokkum}, {Whitaker}, {Brammer}, {Franx}, {Kriek}, {Labb{\'e}}, {Marchesini}, {Quadri}, {Bezanson}, {Illingworth}, {Muzzin}, {Rudnick}, {Tal}, \& {Wake}}]{Dokkum2010}
{van Dokkum}, P.~G., {Whitaker}, K.~E., {Brammer}, G., {et~al.} 2010, \apj, 709, 1018, \dodoi{10.1088/0004-637X/709/2/1018}

\bibitem[{{van Dokkum} {et~al.}(2015){van Dokkum}, {Nelson}, {Franx}, {Oesch}, {Momcheva}, {Brammer}, {F{\"o}rster Schreiber}, {Skelton}, {Whitaker}, {van der Wel}, {Bezanson}, {Fumagalli}, {Illingworth}, {Kriek}, {Leja}, \& {Wuyts}}]{Dokkum2015}
{van Dokkum}, P.~G., {Nelson}, E.~J., {Franx}, M., {et~al.} 2015, \apj, 813, 23, \dodoi{10.1088/0004-637X/813/1/23}

\bibitem[{{Wang} {et~al.}(2024{\natexlab{a}}){Wang}, {de Graaff}, {Davies}, {Greene}, {Leja}, {Goulding}, {Williams}, {Brammer}, {Suess}, {Weibel}, {Bezanson}, {Boogaard}, {Cleri}, {Hirschmann}, {Katz}, {Labbe}, {Maseda}, {Matthee}, {McConachie}, {Naidu}, {Oesch}, {Rix}, {Setton}, \& {Whitaker}}]{Wang2024a}
{Wang}, B., {de Graaff}, A., {Davies}, R.~L., {et~al.} 2024{\natexlab{a}}, arXiv e-prints, arXiv:2403.02304, \dodoi{10.48550/arXiv.2403.02304}

\bibitem[{{Wang} {et~al.}(2024{\natexlab{b}}){Wang}, {Leja}, {de Graaff}, {Brammer}, {Weibel}, {van Dokkum}, {Baggen}, {Suess}, {Greene}, {Bezanson}, {Cleri}, {Hirschmann}, {Labb{\'e}}, {Matthee}, {McConachie}, {Naidu}, {Nelson}, {Oesch}, {Setton}, \& {Williams}}]{Wang2024_balmer}
{Wang}, B., {Leja}, J., {de Graaff}, A., {et~al.} 2024{\natexlab{b}}, \apjl, 969, L13, \dodoi{10.3847/2041-8213/ad55f7}

\bibitem[{{Wang} {et~al.}(2024{\natexlab{c}}){Wang}, {Sun}, {Zhou}, {Xu}, {Cheng}, {Li}, {Chen}, {Mo}, {Dekel}, {Zheng}, {Cai}, {Yang}, {Dai}, {Elbaz}, \& {Huang}}]{Wang2024_mirimasses}
{Wang}, T., {Sun}, H., {Zhou}, L., {et~al.} 2024{\natexlab{c}}, arXiv e-prints, arXiv:2403.02399, \dodoi{10.48550/arXiv.2403.02399}

\bibitem[{{Weaver} {et~al.}(2024){Weaver}, {Cutler}, {Pan}, {Whitaker}, {Labb{\'e}}, {Price}, {Bezanson}, {Brammer}, {Marchesini}, {Leja}, {Wang}, {Furtak}, {Zitrin}, {Atek}, {Chemerynska}, {Coe}, {Dayal}, {van Dokkum}, {Feldmann}, {F{\"o}rster Schreiber}, {Franx}, {Fujimoto}, {Fudamoto}, {Glazebrook}, {de Graaff}, {Greene}, {Juneau}, {Kassin}, {Kriek}, {Khullar}, {Maseda}, {Mowla}, {Muzzin}, {Nanayakkara}, {Nelson}, {Oesch}, {Pacifici}, {Papovich}, {Setton}, {Shapley}, {Shipley}, {Smit}, {Stefanon}, {Taylor}, {Weibel}, \& {Williams}}]{Weaver2024}
{Weaver}, J.~R., {Cutler}, S.~E., {Pan}, R., {et~al.} 2024, \apjs, 270, 7, \dodoi{10.3847/1538-4365/ad07e0}

\bibitem[{{Weibel} {et~al.}(2024){Weibel}, {Oesch}, {Barrufet}, {Gottumukkala}, {Ellis}, {Santini}, {Weaver}, {Allen}, {Bouwens}, {Bowler}, {Brammer}, {Carnall}, {Cullen}, {Dayal}, {Dickinson}, {Donnan}, {Dunlop}, {Giavalisco}, {Grogin}, {Illingworth}, {Koekemoer}, {Labbe}, {Marchesini}, {McLeod}, {McLure}, {Naidu}, {P{\'e}rez-Gonz{\'a}lez}, {Shuntov}, {Stefanon}, {Toft}, \& {Xiao}}]{Weibel2024}
{Weibel}, A., {Oesch}, P.~A., {Barrufet}, L., {et~al.} 2024, \mnras, 533, 1808, \dodoi{10.1093/mnras/stae1891}

\bibitem[{{Weiner} {et~al.}(2006){Weiner}, {Willmer}, {Faber}, {Melbourne}, {Kassin}, {Phillips}, {Harker}, {Metevier}, {Vogt}, \& {Koo}}]{Weiner2006}
{Weiner}, B.~J., {Willmer}, C. N.~A., {Faber}, S.~M., {et~al.} 2006, \apj, 653, 1027, \dodoi{10.1086/508921}

\bibitem[{{Williams} {et~al.}(2024){Williams}, {Alberts}, {Ji}, {Hainline}, {Lyu}, {Rieke}, {Endsley}, {Suess}, {Sun}, {Johnson}, {Florian}, {Shivaei}, {Rujopakarn}, {Baker}, {Bhatawdekar}, {Boyett}, {Bunker}, {Cameron}, {Carniani}, {Charlot}, {Curtis-Lake}, {DeCoursey}, {de Graaff}, {Egami}, {Eisenstein}, {Gibson}, {Hausen}, {Helton}, {Maiolino}, {Maseda}, {Nelson}, {P{\'e}rez-Gonz{\'a}lez}, {Rieke}, {Robertson}, {Saxena}, {Tacchella}, {Willmer}, \& {Willott}}]{Williams2024}
{Williams}, C.~C., {Alberts}, S., {Ji}, Z., {et~al.} 2024, \apj, 968, 34, \dodoi{10.3847/1538-4357/ad3f17}

\bibitem[{{Williams} {et~al.}(2023){Williams}, {Kelly}, {Chen}, {Brammer}, {Zitrin}, {Treu}, {Scarlata}, {Koekemoer}, {Oguri}, {Lin}, {Diego}, {Nonino}, {Hjorth}, {Langeroodi}, {Broadhurst}, {Rogers}, {Perez-Fournon}, {Foley}, {Jha}, {Filippenko}, {Strolger}, {Pierel}, {Poidevin}, \& {Yang}}]{Williams2023_science}
{Williams}, H., {Kelly}, P.~L., {Chen}, W., {et~al.} 2023, Science, 380, 416, \dodoi{10.1126/science.adf5307}

\bibitem[{{Worthey} {et~al.}(1994){Worthey}, {Faber}, {Gonzalez}, \& {Burstein}}]{Worthey1994}
{Worthey}, G., {Faber}, S.~M., {Gonzalez}, J.~J., \& {Burstein}, D. 1994, \apjs, 94, 687, \dodoi{10.1086/192087}

\bibitem[{{Wright} {et~al.}(2024){Wright}, {Whitaker}, {Weaver}, {Cutler}, {Wang}, {Carnall}, {Suess}, {Bezanson}, {Nelson}, {Miller}, {Ito}, \& {Valentino}}]{Wright2024}
{Wright}, L., {Whitaker}, K.~E., {Weaver}, J.~R., {et~al.} 2024, \apjl, 964, L10, \dodoi{10.3847/2041-8213/ad2b6d}

\bibitem[{{Yang} {et~al.}(2022){Yang}, {Morishita}, {Leethochawalit}, {Castellano}, {Calabr{\`o}}, {Treu}, {Bonchi}, {Fontana}, {Mason}, {Merlin}, {Paris}, {Trenti}, {Roberts-Borsani}, {Bradac}, {Vanzella}, {Vulcani}, {Marchesini}, {Ding}, {Nanayakkara}, {Birrer}, {Glazebrook}, {Jones}, {Boyett}, {Santini}, {Strait}, \& {Wang}}]{Yang2022}
{Yang}, L., {Morishita}, T., {Leethochawalit}, N., {et~al.} 2022, \apjl, 938, L17, \dodoi{10.3847/2041-8213/ac8803}

\bibitem[{{Yue} {et~al.}(2024){Yue}, {Eilers}, {Ananna}, {Panagiotou}, {Kara}, \& {Miyaji}}]{Yue2024}
{Yue}, M., {Eilers}, A.-C., {Ananna}, T.~T., {et~al.} 2024, \apjl, 974, L26, \dodoi{10.3847/2041-8213/ad7eba}

\bibitem[{{Yung} {et~al.}(2024){Yung}, {Somerville}, {Finkelstein}, {Wilkins}, \& {Gardner}}]{Yung2024}
{Yung}, L.~Y.~A., {Somerville}, R.~S., {Finkelstein}, S.~L., {Wilkins}, S.~M., \& {Gardner}, J.~P. 2024, \mnras, 527, 5929, \dodoi{10.1093/mnras/stad3484}

\bibitem[{{Zhang} {et~al.}(2015){Zhang}, {Peng}, {C{\^o}t{\'e}}, {Liu}, {Ferrarese}, {Cuillandre}, {Caldwell}, {Gwyn}, {Jord{\'a}n}, {Lan{\c{c}}on}, {Li}, {Mu{\~n}oz}, {Puzia}, {Bekki}, {Blakeslee}, {Boselli}, {Drinkwater}, {Duc}, {Durrell}, {Emsellem}, {Firth}, \& {S{\'a}nchez-Janssen}}]{Zhang2015}
{Zhang}, H.-X., {Peng}, E.~W., {C{\^o}t{\'e}}, P., {et~al.} 2015, \apj, 802, 30, \dodoi{10.1088/0004-637X/802/1/30}

\end{thebibliography}
\bibliographystyle{aasjournal}
\end{document}